\documentclass[twocolumn]{aastex63}
\usepackage{times}
\usepackage{amsmath}
\usepackage{textcomp}
\usepackage{amssymb}
\usepackage{natbib}
\usepackage{float}
\usepackage{color}
\usepackage{graphicx}
\usepackage{epstopdf}
\newcommand{\lum}{erg\,s$^{-1}$}
\newcommand{\fermi}{{\it Fermi}}

\newcommand{\swift}{{\it Swift}}

\newcommand{\phflux}{\mbox{${\rm \, ph \,\, cm^{-2} \, s^{-1}}$}}

\newcommand{\gm}{$\gamma$}

\newcommand{\dcf}{$z$-DCF}
\newcommand{\tsv}{TS$_{\rm var}$}

\submitjournal{ApJ}

\shorttitle{VHE Flux Variability of \fermi~Blazars}
\shortauthors{Vaidehi S. Paliya}

\begin{document}
\title{Very High-Energy ($>$50 GeV) Gamma-ray Flux Variability of Bright Fermi Blazars}

\correspondingauthor{Vaidehi S. Paliya}

\author[0000-0001-7774-5308]{Vaidehi S. Paliya}
\affiliation{Inter-University Centre for Astronomy and Astrophysics (IUCAA), SPPU Campus, 411007, Pune, India}

\email{vaidehi.s.paliya@gmail.com}

\begin{abstract}
Understanding the high-energy emission processes and variability patterns are two of the most challenging research problems associated with relativistic jets. In particular, the long-term (months-to-years) flux variability at very high energies (VHE, $>$50 GeV) has remained an unexplored domain so far. This is possibly due to the decreased sensitivity of the \fermi~Large Area Telescope (LAT) above a few GeV, hence low photon statistics, and observing constraints associated with the ground-based Cherenkov telescopes. This paper reports the results obtained from the 0.05$-$2 TeV \fermi-LAT data analysis of a sample of 29 blazars with the primary objective to explore their months-to-year long VHE flux variability behavior. This systematic search has led to, for the first time, the detection of significant flux variations in 5 blazars at $>$99\% confidence level, whereas, 8 of them exhibit variability albeit at a lower confidence level ($\sim$95\%-99\%). A comparison of the 0.05$-$2 TeV flux variations with that observed at 0.1$-$50 GeV band has revealed similar variability behavior for most of the sources. However, complex variability patterns that are not reflected contemporaneously in both energy bands were also detected, thereby providing tantalizing clues about the underlying radiative mechanisms. These results open up a new dimension to unravel the VHE emission processes operating in relativistic jets, hence sowing the seeds for their future observations with the upcoming Cherenkov Telescope Array.

\end{abstract}


\section{Introduction}{\label{sec:Intro}}
The high-energy emission and the erratic, rapid, and large amplitude variability observed in all accessible spectral regimes (radio-to-\gm-ray) are two of the main defining properties of blazars \citep[e.g.,][]{2016Galax...4...37M}, a class of radio-loud active galactic nuclei (AGN) that have powerful relativistic jets oriented close to the line of sight to the observer. They are the most variable (on years-to-minute timescales), the most luminous ($L_{\rm bol.}>10^{46}$ \lum), the most polarized (even $>30$\% in the optical and radio bands), and thus probably the most exciting type of AGN. The emitted radiation from the jet is strongly amplified due to relativistic effects, so-called Doppler boosting. This flux enhancement makes blazars one of the very few types of astrophysical objects detectable throughout the accessible electromagnetic spectrum and the most abundant sources in the extragalactic \gm-ray sky \citep[e.g.,][]{2022ApJS..263...24A}. Blazars are categorized into flat spectrum radio quasars (FSRQs) and BL Lac objects based on the rest-frame equivalent width (EW) of their broad optical emission lines, with BL Lac sources having EW$<$5 \AA~\citep[cf.][]{1991ApJ...374..431S}.
\vskip 0.1cm
The spectral energy distribution (SED) of a blazar is dominated by non-thermal jet emission and characterized by two broad humps. The low energy peak, produced by the synchrotron process, lies in the sub-millimeter (mm) to the UV/X-ray energy regime, whereas, the high energy hump from inverse Compton mechanism peaks in the MeV/GeV range, assuming leptons to be responsible for the observed radiation. Alternatively, hadronic processes ($\pi^0$ decay and subsequent cascade radiation) have also been proposed to explain the high-energy emission from blazars \citep[e.g.,][]{2013ApJ...768...54B}. Interestingly, hadronic models predict blazar jets as promising candidates for neutrino emission \citep[cf.][]{1989A&A...221..211M,2019ApJ...881...46R}. This indicates that relativistic jets can be among the most efficient particle accelerators. Furthermore, based on the location of the synchrotron peak frequency ($\nu^{\rm peak}_{\rm syn}$), blazars are also classified as low- ($\nu^{\rm peak}_{\rm syn}<$10$^{14}$ Hz), intermediate- (10$^{14}<\nu^{\rm peak}_{\rm syn}<$10$^{15}$ Hz) and high- ($\nu^{\rm peak}_{\rm syn}>$10$^{15}$ Hz) synchrotron peaked sources \citep[][]{2010ApJ...716...30A}. 

Flux variability is probably the most studied topic in jet physics, yet we do not have a clear consensus about the underlying physical processes, the behavior of the variability pattern across various blazar sub-classes and connection with the accretion and central black hole, the variability timescales, and multi-band flux correlations. There have been several studies done to understand the blazar flux variability across the electromagnetic spectrum, e.g., at radio \citep[cf.][]{2016A&A...596A..45F}, infrared \citep[e.g.,][]{2018Ap&SS.363..167M,2020MNRAS.494..764A},  optical-ultraviolet \citep[cf.][]{1992ApJ...401..516E,2004JApA...25....1S,2012ApJ...756...13B,2017ApJ...844...32P}, X-ray \citep[][]{2002PASA...19...49P,2017MNRAS.466.3309R}, and \gm-ray energies \citep[cf.][]{2010ApJ...722..520A}. Additionally, multi-wavelength studies focused both on short- ($\lesssim$days) and long-term (months-to-years) flux variations have been carried out to understand the radiative processes, jet composition, and kinematics \citep[e.g.,][]{1998ApJ...501L..51B,2005ApJ...629...52S,2011MNRAS.413..333N,2014ApJ...780...87M,2015ApJ...814...51T,2019ApJ...887..133B}. 

At very high energies (VHE, $>$50 GeV), blazar observations are often triggered by elevated activity states identified at lower frequencies \citep[e.g.,][]{2022ApJ...932..129A}. These follow-up observations with Cherenkov telescopes usually cover the source activity for days to weeks, thus providing an opportunity to study the event on short timescales. The long-term, i.e., months-to-years, VHE flux variations of blazars, on the other hand, remained almost an unexplored territory with studies have been limited so for to the brightest sources \citep[e.g.,][]{2014APh....54....1A,2017ApJ...841..100A,2020ApJ...891..170V}. This is possibly due to telescope time availability, weather constraints, and/or source visibility from the ground-based facilities. This work attempts to address this outstanding research problem using VHE observations carried out with the \fermi-Large Area Telescope \citep[\fermi-LAT,][]{2009ApJ...697.1071A}. Due to the surveying capabilities of the \fermi-LAT, almost uninterrupted monitoring of the whole \gm-ray sky is possible at energies as high as $\gtrsim$2 TeV. Though the sensitivity of the \fermi-LAT starts decreasing above a few GeV energies,  so the background contamination. This enables an improved source localization and, thus, the more accurate association of the \gm-ray photons with a point source. Admittedly, the temporal VHE flux variations cannot be studied using the \fermi-LAT on shorter timescales due to the faintness of blazars at these energies, hence low photon statistics. However, integrating the LAT data on longer timescales allows one to study the months-to-year flux variations. This is the primary objective of this work. In Section~\ref{sec2}, and~\ref{sec3}, the details of the sample selection and data reduction steps are described.  The derived results are presented in Section~\ref{sec4}, and noteworthy points about individual blazars are highlighted in Section~\ref{sec5}. The overall findings are summarized in Section~\ref{sec6}.

\section{Sample Selection}\label{sec2}
The third data release of the fourth catalog of \fermi-LAT detected \gm-ray sources \citep[4FGL-DR3,][]{2022ApJS..260...53A} includes 172 objects with detection significance $\gtrsim$4.2$\sigma$ \citep[test statistic TS $>$25,][]{1996ApJ...461..396M} in the energy range of 100 GeV-to-1 TeV. Considering only blazars or blazar candidates of uncertain type (BCU), the sample size was reduced to 125 sources. The \fermi-LAT data reduction covering the first $\sim$15 years of the mission was carried out for all of them using the methodology described in Section~\ref{sec3}. Only those 29 objects were retained whose overall detection significance was estimated to be $\gtrsim$10$\sigma$ (TS $>$100). This criterion ensures that a meaningful temporal study can be done on the selected \gm-ray sources.  This final sample contains 28 BL Lac objects and one BCU listed in Table~\ref{tab:basic_info}. All sources have been flagged as variable in the 4FGL-DR3 catalog \citep[see also][]{2023ApJS..265...31A}.

\begin{table*}
\caption{The spectral information of blazars in 0.1$-$2 TeV energy range derived from the analysis of $\sim$15 years of the \fermi-LAT observations. Column information are as follows: 
(1) 4FGL name; (2) association; (3) redshift; (4) (5), and (6) observed photon flux (in units of 10$^{-11}$ \phflux), photon index, and test statistic, in 0.1$-$2 TeV energy range; (7) (8), and (9) EBL-corrected photon flux (in units of 10$^{-11}$ \phflux), photon index, and test statistic, in 0.1$-$2 TeV energy range; (10) whether the source was detected with ground-based Cherenkov telescopes, Y: yes, N: no; (11) variability test statistic; and (12) status whether a source is variable (V), probably variable (PV), or non-variable (NV) in 0.1$-$2 TeV energy range.}\label{tab:basic_info}
\begin{center}
\begin{tabular}{llllllllllll}
\hline
4FGL name  & Other name & $z$ & $F_{\rm obs}$ & $\Gamma_{\rm obs}$ &  TS$_{\rm obs}$  & $F_{\rm EBL}$ & $\Gamma_{\rm EBL}$ &  TS$_{\rm EBL}$ & TeV & TS$_{\rm var}$ & Status\\
(1) & (2) & (3)  & (4) & (5) &  (6) & (7) & (8) & (9) & (10) & (11) & (12)\\ 

\hline
J0035.9+5950     & 1ES 0033+595            & 0.47 &  4.19$\pm$0.70  & 2.92$\pm$0.35  & 326 & 4.25$\pm$0.71  & 1.03$\pm$0.44 & 328 & Y & 67.1 & NV\\
J0222.6+4302     & 3C 66A                       & 0.44 &  4.00$\pm$0.76  & 3.76$\pm$0.53  & 226 & 4.06$\pm$0.76  & 2.39$\pm$0.63 & 228 & Y & 80.4 & PV\\
J0449.4$-$4350     & PKS 0447$-$439       & 0.21 &  5.61$\pm$0.88  & 3.20$\pm$0.37  & 445 & 5.65$\pm$0.89  & 2.57$\pm$0.41 & 446 & Y & 85.3 & PV\\
J0507.9+6737      &  1ES 0502+675          & 0.42 &  4.32$\pm$0.69  & 2.88$\pm$0.33  & 418 & 4.36$\pm$0.69  & 1.26$\pm$0.41 & 416 & Y & 115.8 & V\\
J0521.7+2112        &  TXS 0518+211        & 0.11 &  4.89$\pm$0.91  & 3.02$\pm$0.41  & 233 & 4.90$\pm$0.91  & 2.71$\pm$0.44 & 232 & Y & 82.3 & PV\\
J0543.9$-$5531  &  RX J0543.9$-$5532     & 0.27 &  2.63$\pm$0.58  & 2.50$\pm$0.41  & 220 & 2.60$\pm$0.59  & 1.50$\pm$0.54 & 221 & N & 16.7 & NV\\
J0648.7+1516       &  RX J0648.7+1516     & 0.18 &  2.00$\pm$0.60  & 1.65$\pm$0.40  & 106 & 1.99$\pm$0.60  & 0.78$\pm$0.44 & 106 & Y & 26.5 & NV\\
J0650.7+2503      &  1ES 0647+250          & 0.20 &  5.54$\pm$0.96  & 3.26$\pm$0.43  & 333 & 5.56$\pm$0.96  & 2.67$\pm$0.48 & 333 & Y & 22.9 & NV\\
J0721.9+7120        & S5 0716+71             & 0.13 &  2.69$\pm$0.54  & 3.59$\pm$0.54  & 216 & 2.70$\pm$0.54  & 3.26$\pm$0.57  & 217 & Y & 87.8 & PV\\
J0809.8+5218      & 1ES 0806+524           & 0.14 &  2.49$\pm$0.57  & 2.85$\pm$0.46  & 198 & 2.50$\pm$0.57  & 2.39$\pm$0.50  & 200 & Y & 48.9 & NV\\
J1015.0+4926        & 1H 1013+498           & 0.21 &  4.47$\pm$0.77  & 2.66$\pm$0.33  & 315 & 4.48$\pm$0.77  & 1.87$\pm$0.37  & 318 & Y & 52.8 & NV\\
J1037.7+5711         & GB6 J1037+5711     &        &  1.07$\pm$0.35  & 4.78$\pm$1.26 & 108 &                          &                          &        & N & 37.0 & NV\\
J1104.4+3812         & Mkn 421                  & 0.03 &  52.80$\pm$2.79& 2.20$\pm$0.08 & 5942& 52.80$\pm$2.79& 2.08$\pm$0.09  & 5943 & Y & 150.7 & V\\
J1120.8+4212        & RBS 0970                 & 0.12  & 1.76$\pm$0.50  & 3.89$\pm$0.83 & 141 & 1.77$\pm$0.50  & 3.60$\pm$0.87   & 141 & N & 46.1 & NV\\
J1136.4+7009        & Mkn 180                   & 0.05  & 1.33$\pm$0.37 & 3.07$\pm$0.62 & 130 & 1.34$\pm$0.37   &  2.94$\pm$0.63  & 131 & Y & 34.9 & NV\\
J1217.9+3007        & 1ES 1215+303          & 0.13  & 2.32$\pm$0.61 & 2.85$\pm$0.53 & 118 & 2.33$\pm$0.61   & 2.42$\pm$0.57   & 118  & Y & 81.1 & PV\\
J1221.3+3010         & PG 1218+304           & 0.18  & 6.16$\pm$0.99 & 2.02$\pm$0.24 & 431 & 6.15$\pm$0.96  & 1.20$\pm$0.25   & 435 & Y & 86.4 & PV\\
J1427.0+2348        & PKS 1424+240          & 0.60  & 4.94$\pm$0.85 & 4.58$\pm$0.62 & 329 & 5.04$\pm$0.86  & 2.87$\pm$0.71   & 331 & Y & 70.2 & NV\\
J1443.9$-$3908        & PKS 1440$-$389     & 0.14  & 2.71$\pm$0.64 & 3.39$\pm$0.59 & 204 & 2.72$\pm$0.64 & 3.23$\pm$0.61    & 204 & Y & 44.2 & NV\\
J1555.7+1111        & PG 1553+113            & 0.43  & 13.50$\pm$1.44&3.64$\pm$0.30 & 863 & 13.60$\pm$1.46&2.57$\pm$0.34    & 869 & Y & 80.3 & PV\\
J1653.8+3945        & Mkn 501                   & 0.03  & 15.50$\pm$1.41&2.54$\pm$0.16 & 1736 & 15.50$\pm$1.41&2.43$\pm$0.17  & 1736 & Y & 135.7 & V\\
J1728.3+5013        & I Zw 187                   & 0.06  & 1.68$\pm$0.46 &1.95$\pm$0.39 & 118 & 1.68$\pm$0.46 &1.73$\pm$0.40     & 118  & Y & 41.5 & NV\\
J1917.7$-$1921        & 1H 1914$-$194       & 0.14  & 2.37$\pm$0.64 & 3.05$\pm$0.59 & 142 & 2.38$\pm$0.64 & 2.63$\pm$0.63   & 142 & N & 42.3 & NV\\
J2000.0+6508        & 1ES 1959+650          & 0.05  & 13.80$\pm$1.20& 2.29$\pm$0.14 & 1701 & 13.80$\pm$1.20& 2.12$\pm$0.15 & 1703 & Y & 115.6 & V\\
J2009.4$-$4849    & PKS 2005$-$489         & 0.07  & 2.96$\pm$0.66 & 2.93$\pm$0.46 & 181 & 2.97$\pm$0.66 & 2.71$\pm$0.48 & 182 & Y & 27.9 & NV \\
J2056.7+4939         & RGB J2056+496       &  0.10 & 2.44$\pm$0.56 & 1.95$\pm$0.33 & 136 & 2.44$\pm$0.56 & 1.51$\pm$0.35 & 138 & Y & 28.6 & NV\\
J2158.8$-$3013      & PKS 2155$-$304       & 0.12  & 13.30$\pm$1.50&2.70$\pm$0.22 & 1044 & 13.30$\pm$1.50&2.30$\pm$0.23 & 1045 & Y & 88.1 & PV\\
J2202.7+4216          & BL Lacertae             & 0.07  & 5.58$\pm$0.86  & 2.81$\pm$0.32& 503 & 5.58$\pm$0.86  & 2.59$\pm$0.33 & 504 & Y & 131.8 & V\\
J2347.0+5141           & 1ES 2344+514       & 0.04  & 4.33$\pm$0.74  & 2.18$\pm$0.27 & 326 & 4.33$\pm$0.74  & 2.01$\pm$0.28 & 327 & Y & 30.8 & NV \\
\hline
\end{tabular}
\end{center}
\end{table*}

\section{Fermi-LAT data reduction}\label{sec3}
The standard data reduction steps were followed to analyze the $\sim$15 years of the \fermi-LAT data (modified Julian date or MJD 54683$-$60220, i.e., 2008 August 5 to 2023 October 2). In the energy range of 0.1$-$2 TeV, only SOURCE class events were selected from a circular region of radius 5$^{\circ}$ which passed the zenith angle-based cut ($z_{\rm max}<105^{\circ}$). The Pass 8 data includes division of photons by point-spread function (PSF) type with the worst (PSF0) and best (PSF3) best angular reconstructions, respectively. Therefore, a component-wise LAT data analysis for each PSF was performed. The final fitting was carried out using the SUMMED likelihood method of the FermiTools. For the likelihood fitting, all \gm-ray sources present in the 4FGL-DR3 catalog and lying within 7$^{\circ}$ from the position of interest were considered to represent the \gm-ray sky. The model file also includes latest diffuse background models\footnote{\url{https://fermi.gsfc.nasa.gov/ssc/data/access/lat/BackgroundModels.html}} \citep[e.g.,][]{2016ApJS..223...26A}. Due to low photon statistics, the \gm-ray spectrum of the source of interest was assumed to follow a power law spectral shape. The first round of the model optimization was performed using fermiPy \citep[][]{2017arXiv170709551W} to estimate the detection significance, i.e., TS, of all sources and the spectral parameters of those with TS $>$25 were left free to vary during the likelihood fitting. TS maps were also generated to search for any unmodeled \gm-ray objects present in the data but not in the model since the time period covered in the analysis is larger than that used to produce the 4FGL-DR3 catalog. Once identified, if any, they were included in the model assuming a power law spectral shape, and the likelihood fit was repeated. Since the extragalactic background light (EBL) attenuation of \gm-ray photons becomes important at the considered energy range, the spectral fitting was also performed including the EBL absorption to determine the EBL-corrected spectral parameters. The EBL model proposed by \citet[][]{2011MNRAS.410.2556D} was used for this purpose.

After the $\sim$15 years averaged data analysis, 29 objects were selected whose overall TS was estimated to be $>$100. The 0.1$-$2 TeV spectral parameters of these sources are provided in Table~\ref{tab:basic_info}. To generate the \gm-ray light curves, 3 months time binning was adopted and the minimum energy considered in the analysis was changed to 50 GeV. This was done to improve the photon statistics while ensuring that the analysis probes the VHE domain which is the main objective of the work. The EBL-corrected spectral parameters, i.e., normalization and photon index, of the source of interest were allowed to vary during the likelihood fitting carried out in the individual time bin. Whenever the TS of the target object was found to be $<$4, the flux upper limit at 95\% confidence level was derived. This choice is slightly different from that conventionally adopted while analyzing 0.1$-$300 GeV data, i.e., TS$<9$. However, all blazars under consideration are well-detected with the \fermi-LAT and several of them have also been detected with the ground-based Cherenkov telescopes (Table~{\ref{tab:basic_info}}). Therefore, a slightly reduced detection significance can be adopted to maximize the number of data points in the light curves.

To compare the flux variations observed in 0.05$-$2 TeV energy range (hereafter TeV band) with that at lower \gm-ray energies, the light curves were also generated in 0.1$-$50 GeV energy band (hereafter GeV band) using the same time binning. An ROI of 15$^{\circ}$ radius was chosen, and all 4FGL-DR3 sources lying within 20$^{\circ}$ from the position of interest were considered for the likelihood fitting. The spectral shapes were directly taken from the 4FGL-DR3 catalog to model the 0.1$-$50 GeV energy spectrum. 

\begin{figure*}
\hbox{
\includegraphics[scale=0.23]{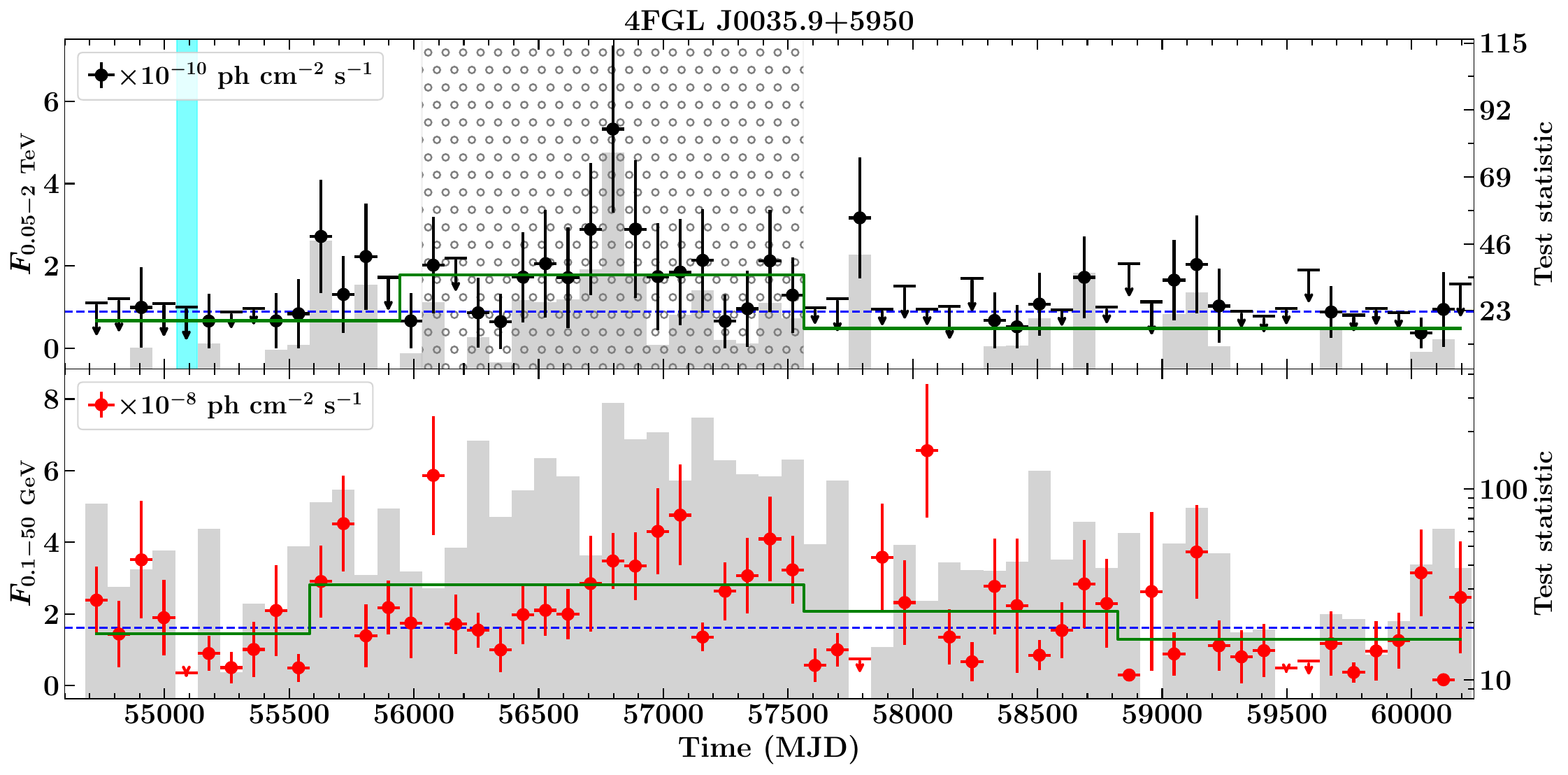}
\includegraphics[scale=0.23]{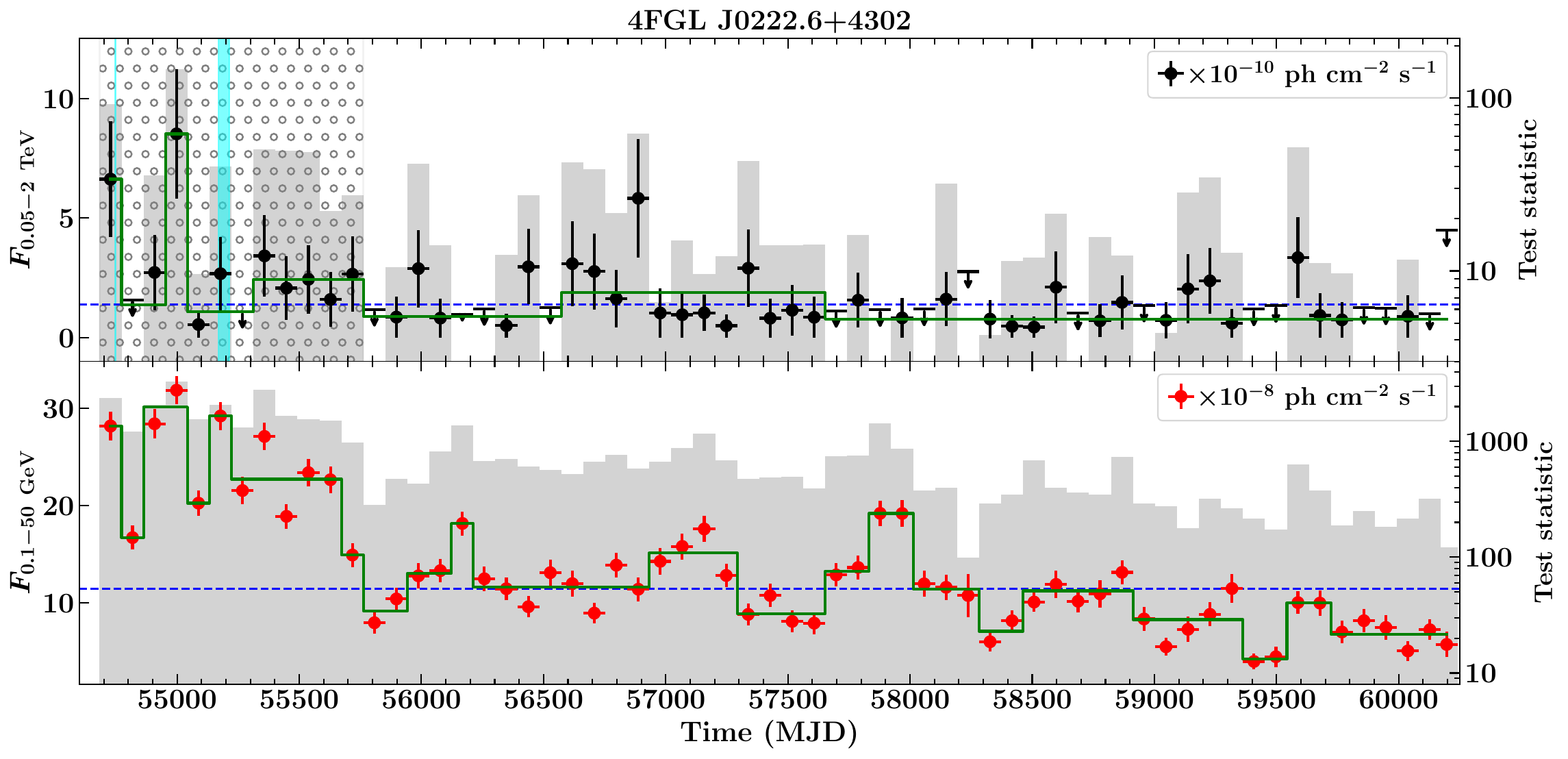}
}
\hbox{
\includegraphics[scale=0.23]{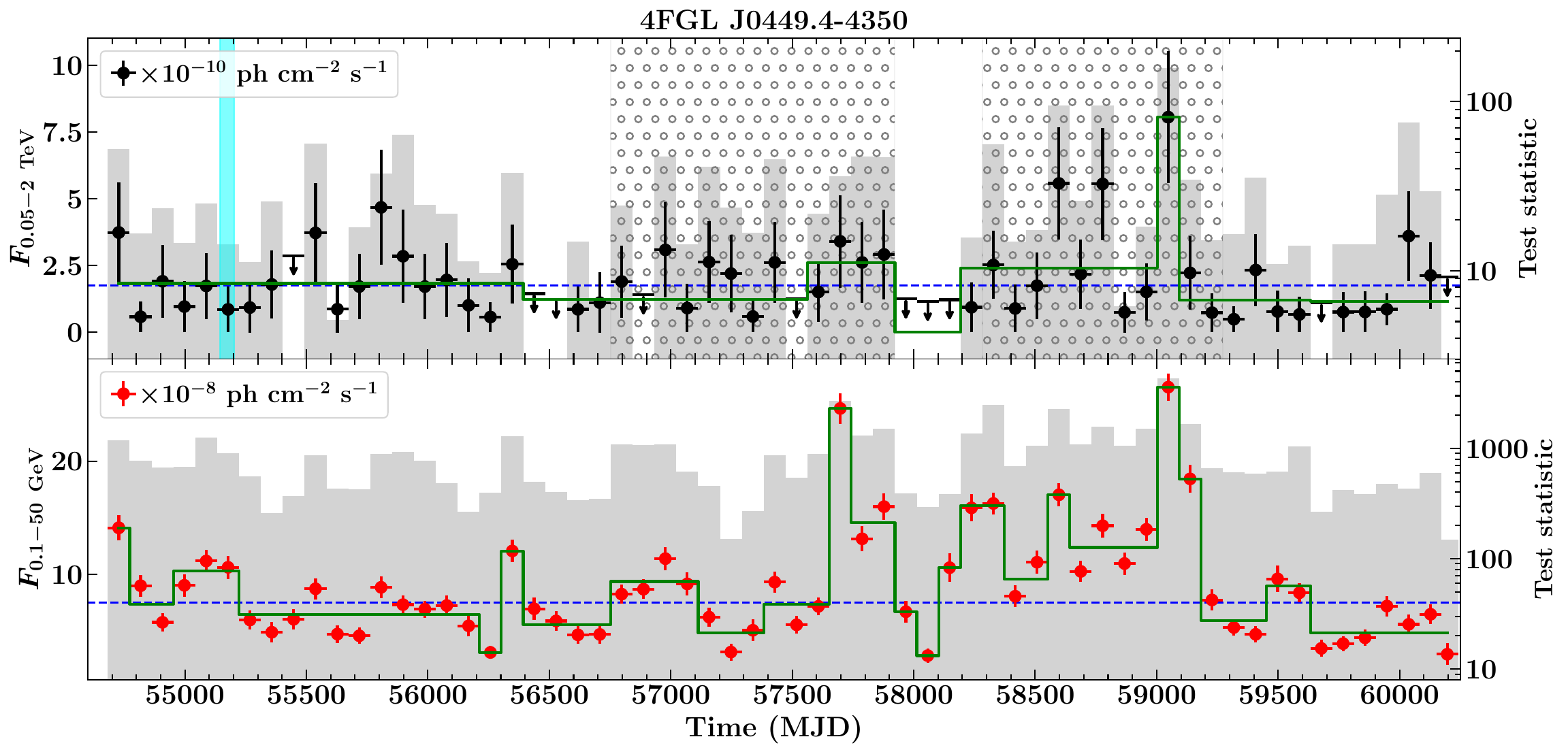}
\includegraphics[scale=0.23]{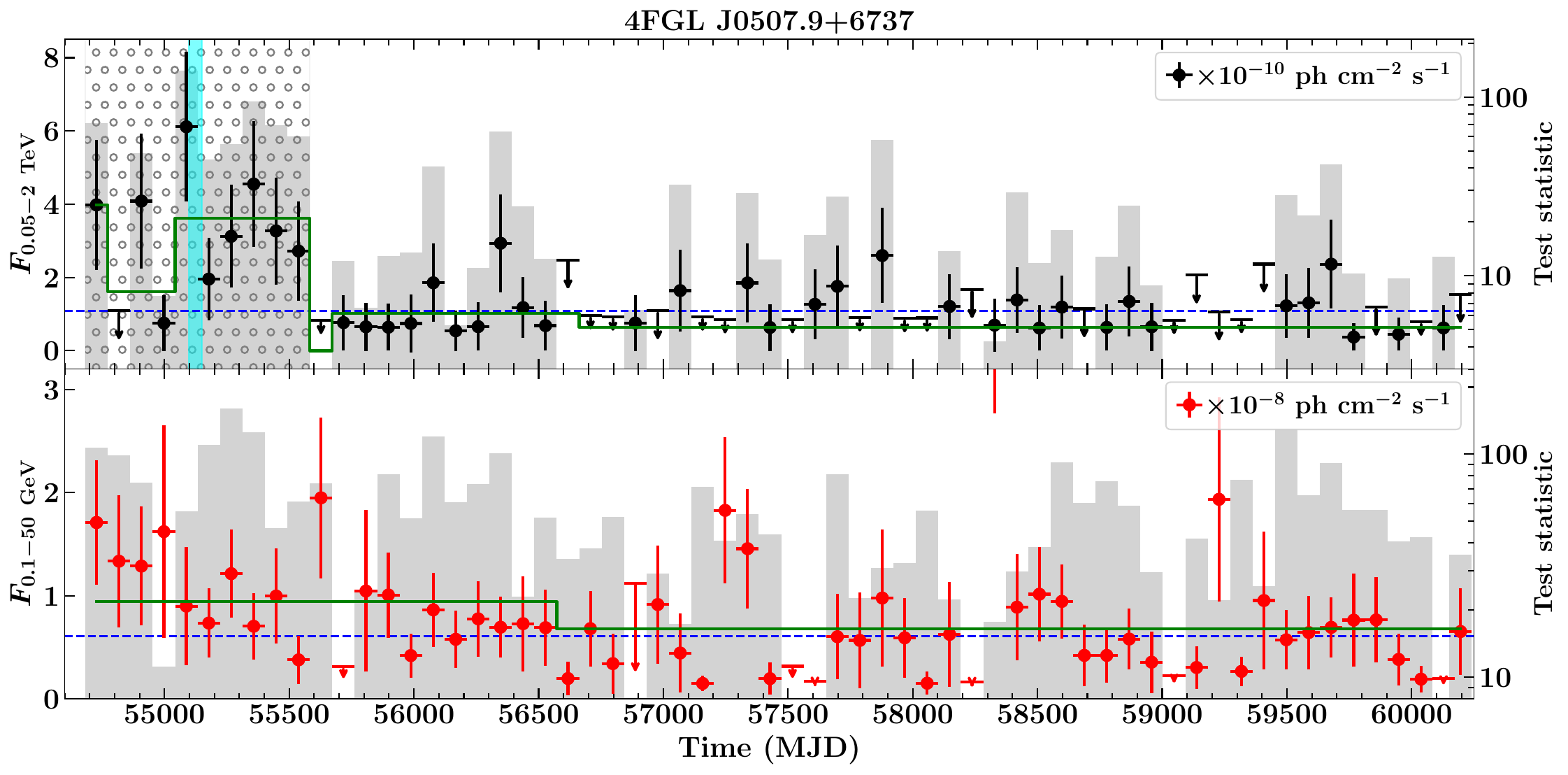}
}
\hbox{
\includegraphics[scale=0.23]{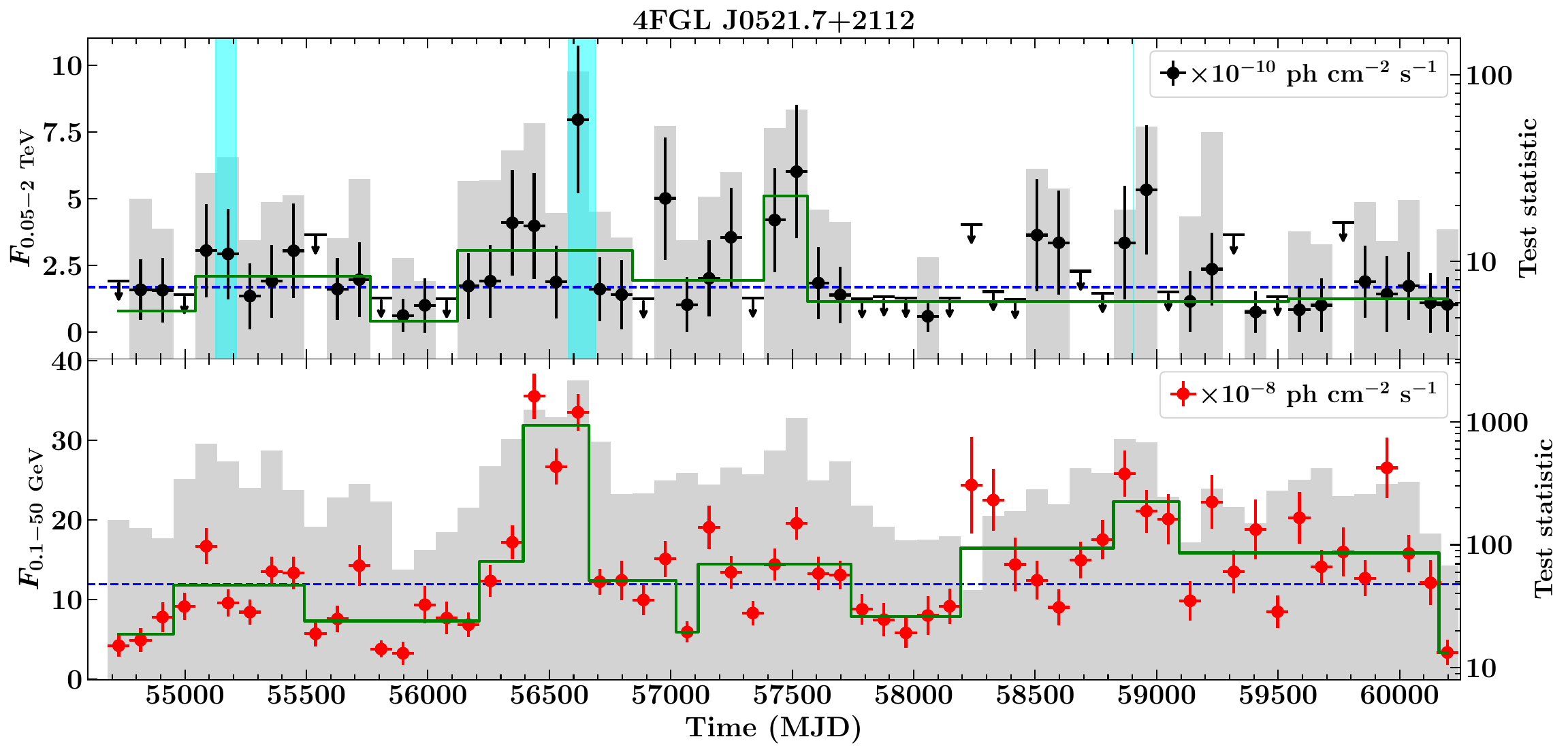}
\includegraphics[scale=0.23]{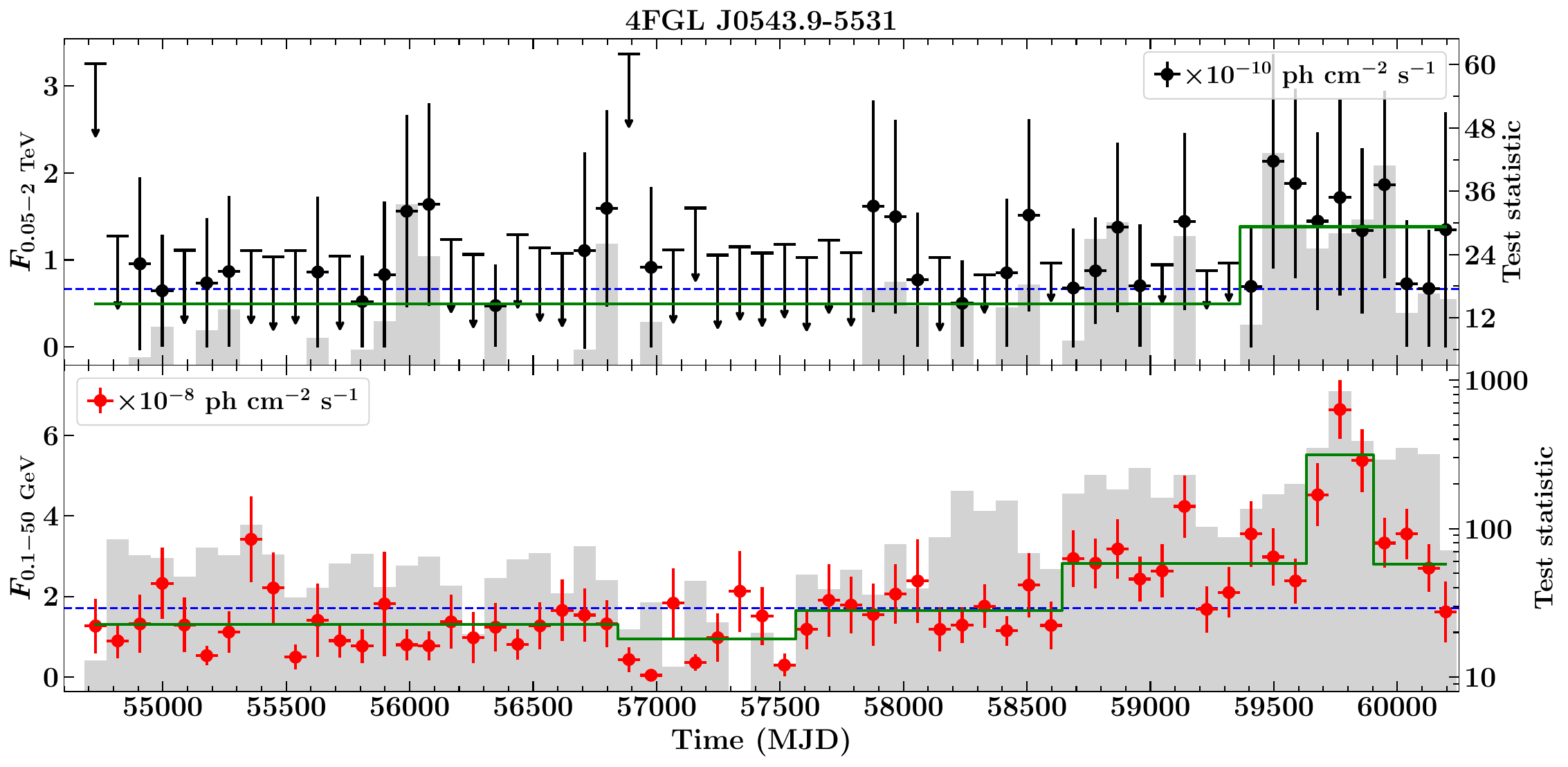}
}
\hbox{
\includegraphics[scale=0.23]{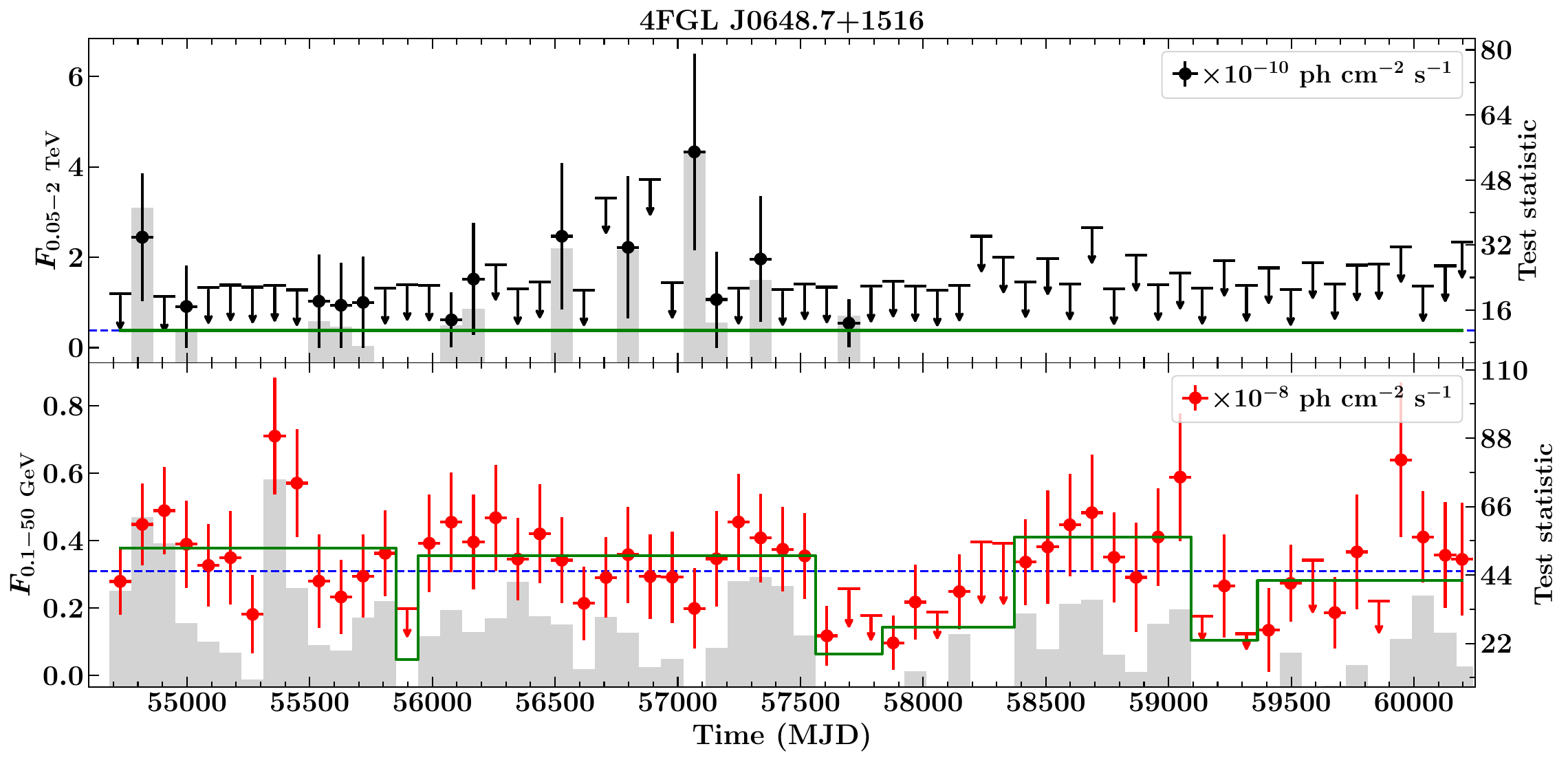}
\includegraphics[scale=0.23]{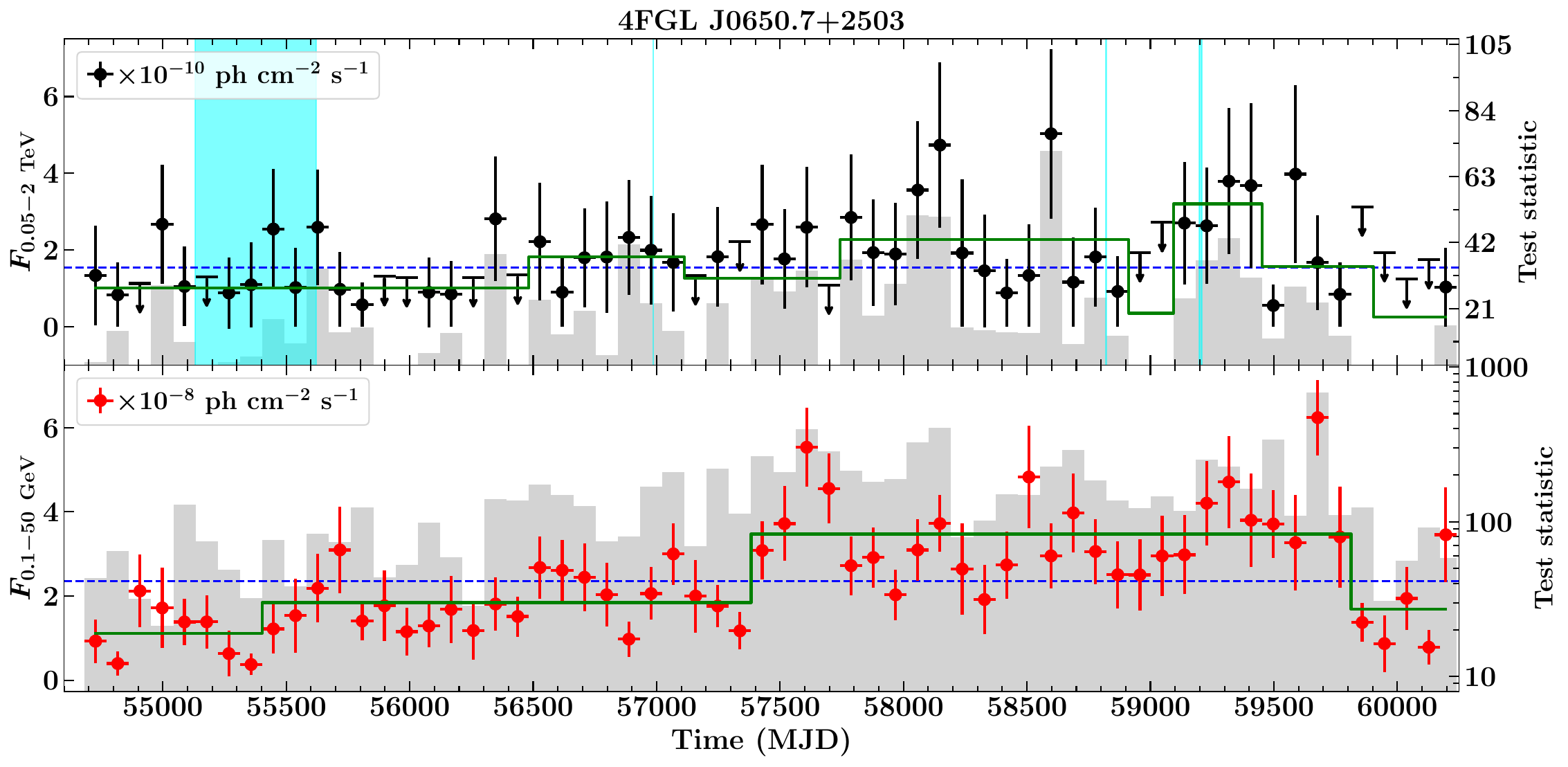}
}
\hbox{
\includegraphics[scale=0.23]{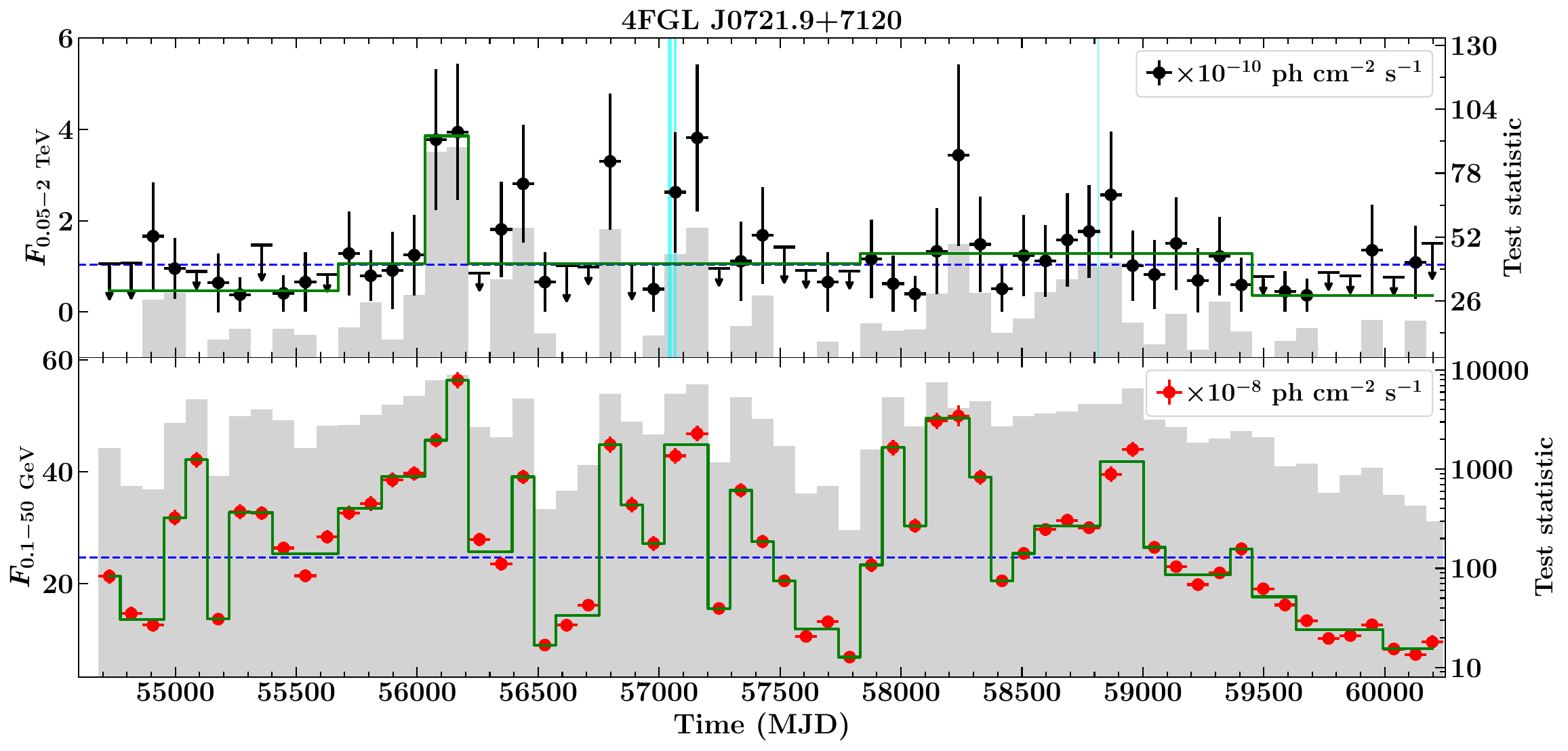}
\includegraphics[scale=0.23]{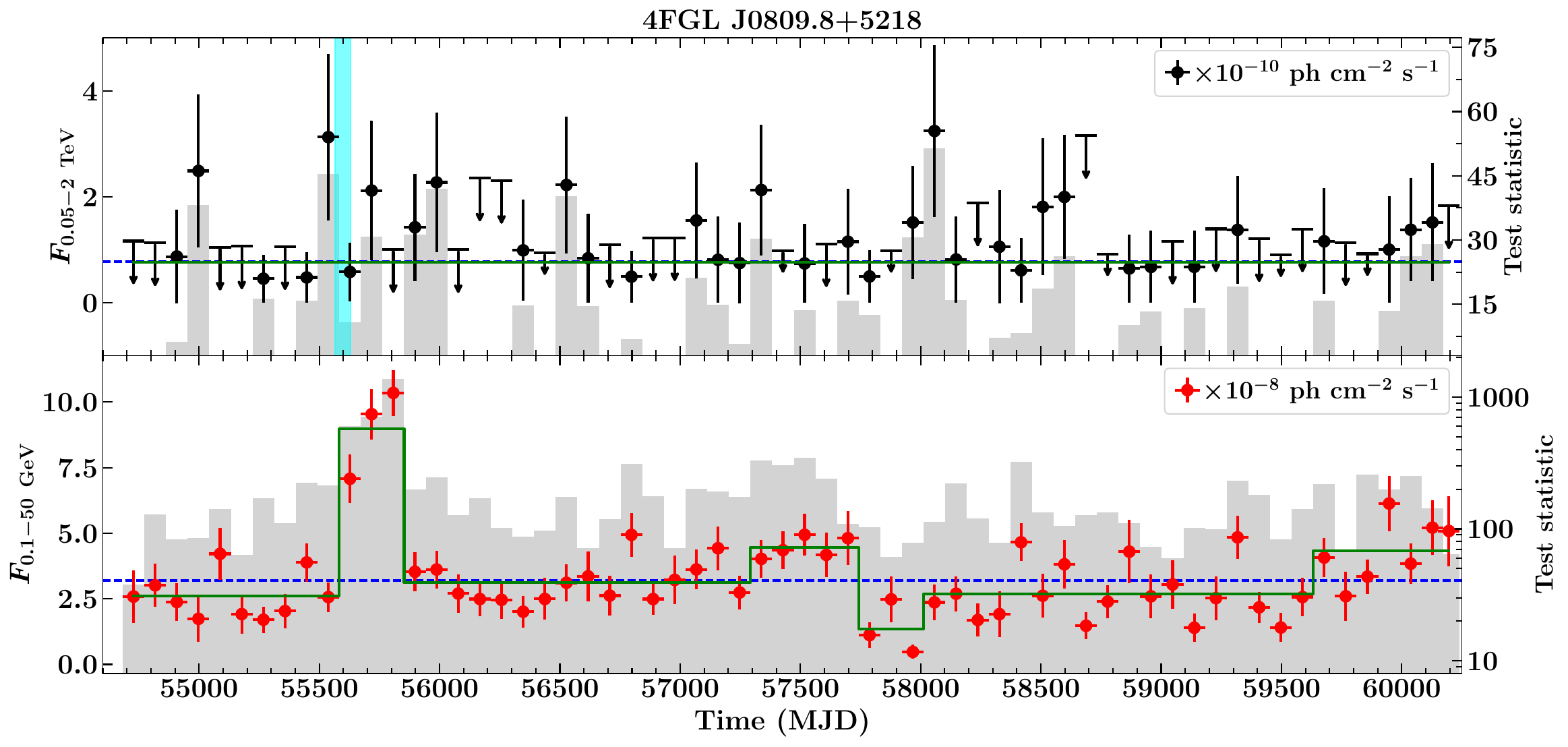}
}

\caption{Three-months binned \gm-ray light curves of VHE detected blazars studied in this work. In each figure, the top panel shows the flux variations in the 0.05$-$2 TeV energy range, whereas, the 0.1$-$50 GeV light curve is plotted in the bottom panel.  Upper limits are derived at 95\% confidence level and plotted with downward arrows. The grey bars represent the TS value in each time bin. In both panels, the blue dashed and the green solid lines represent the $\sim$15 years averaged \gm-ray flux and Bayesian blocks, respectively, for the source under consideration. The vertical cyan lines highlight the periods of detection with Cherenkov telescopes. The hatched areas refer to the periods for which the \dcf~analysis was carried out.}\label{fig:1}
\end{figure*}

\begin{figure*}
\figurenum{1}
\hbox{
\includegraphics[scale=0.23]{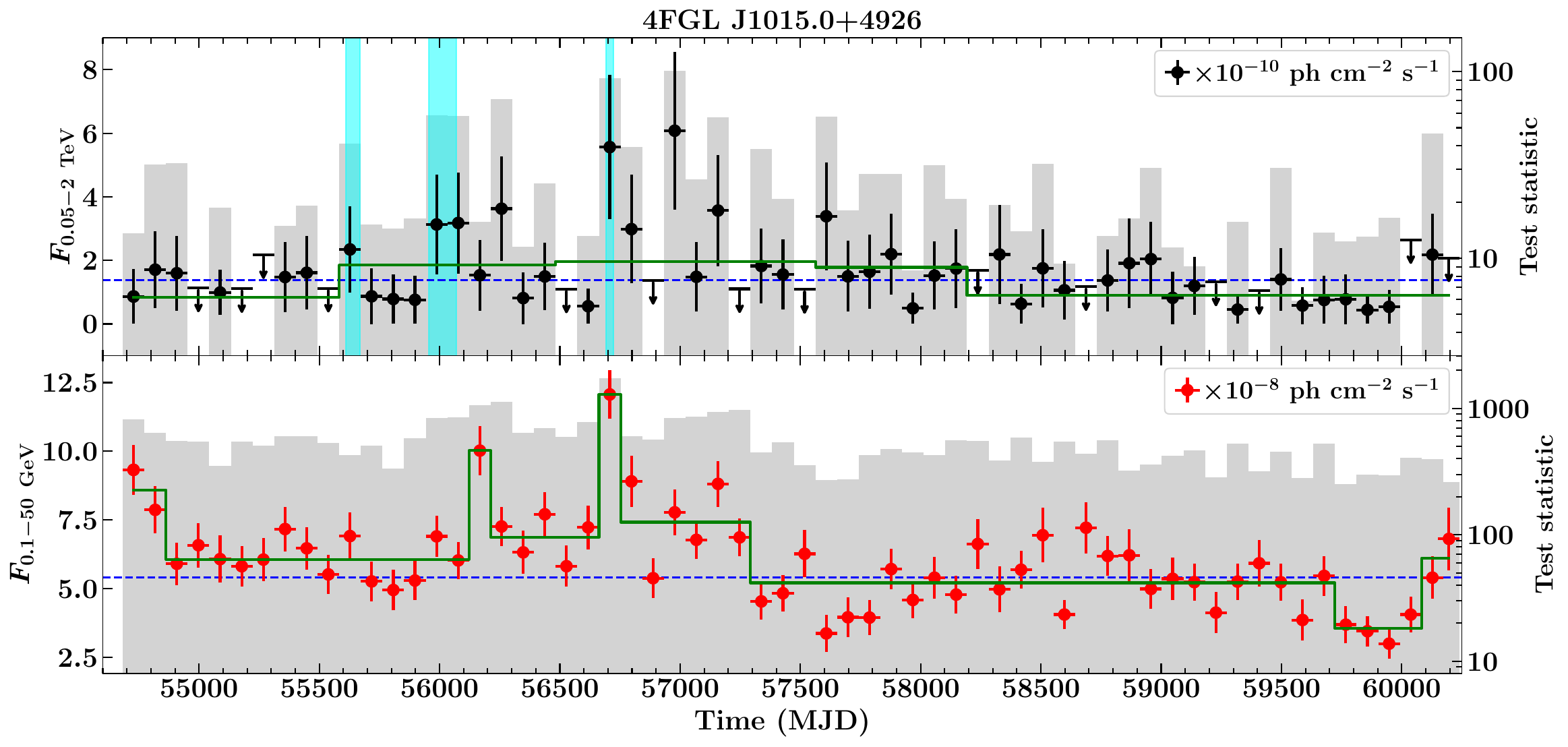}
\includegraphics[scale=0.23]{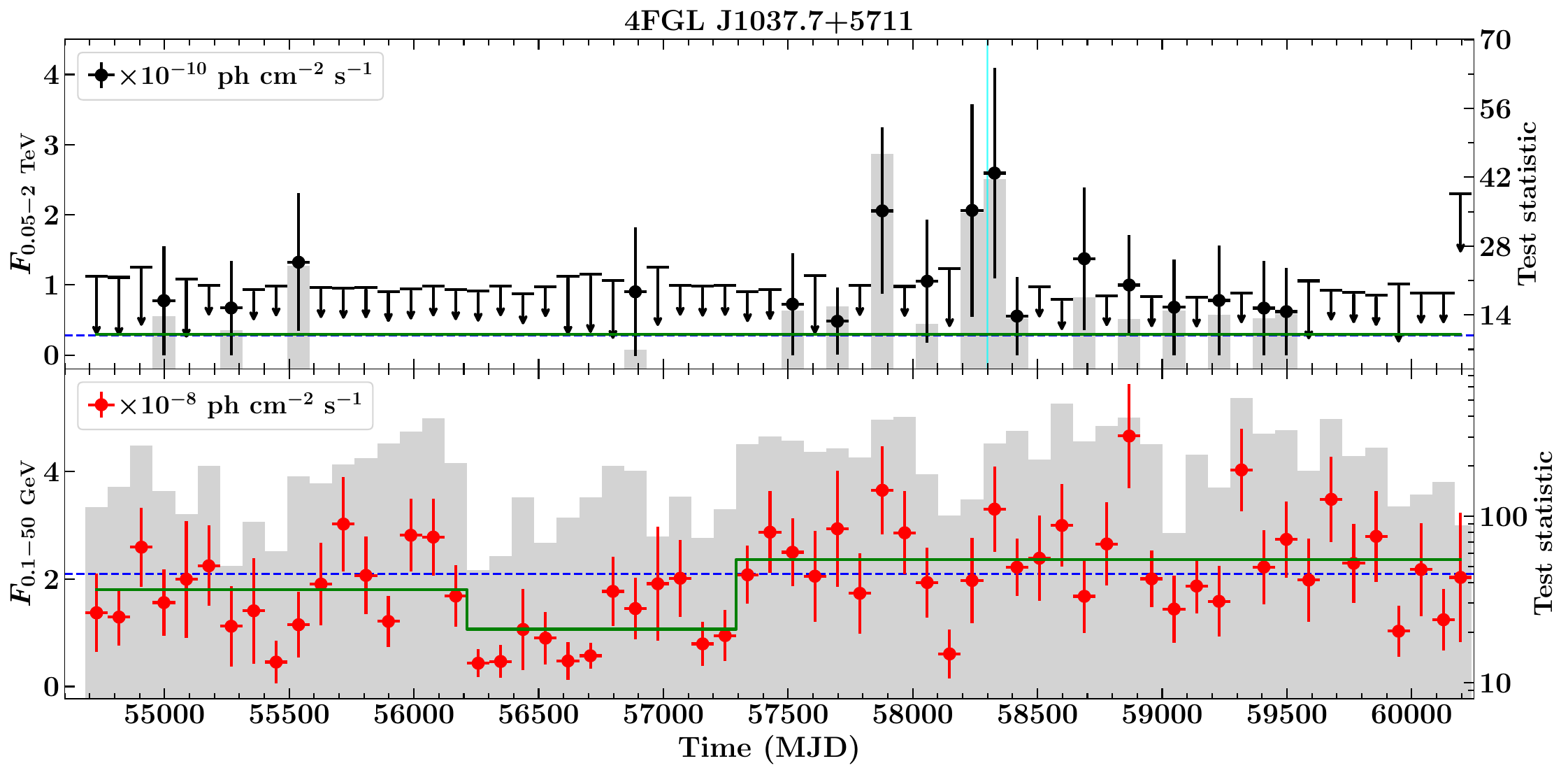}
}
\hbox{
\includegraphics[scale=0.23]{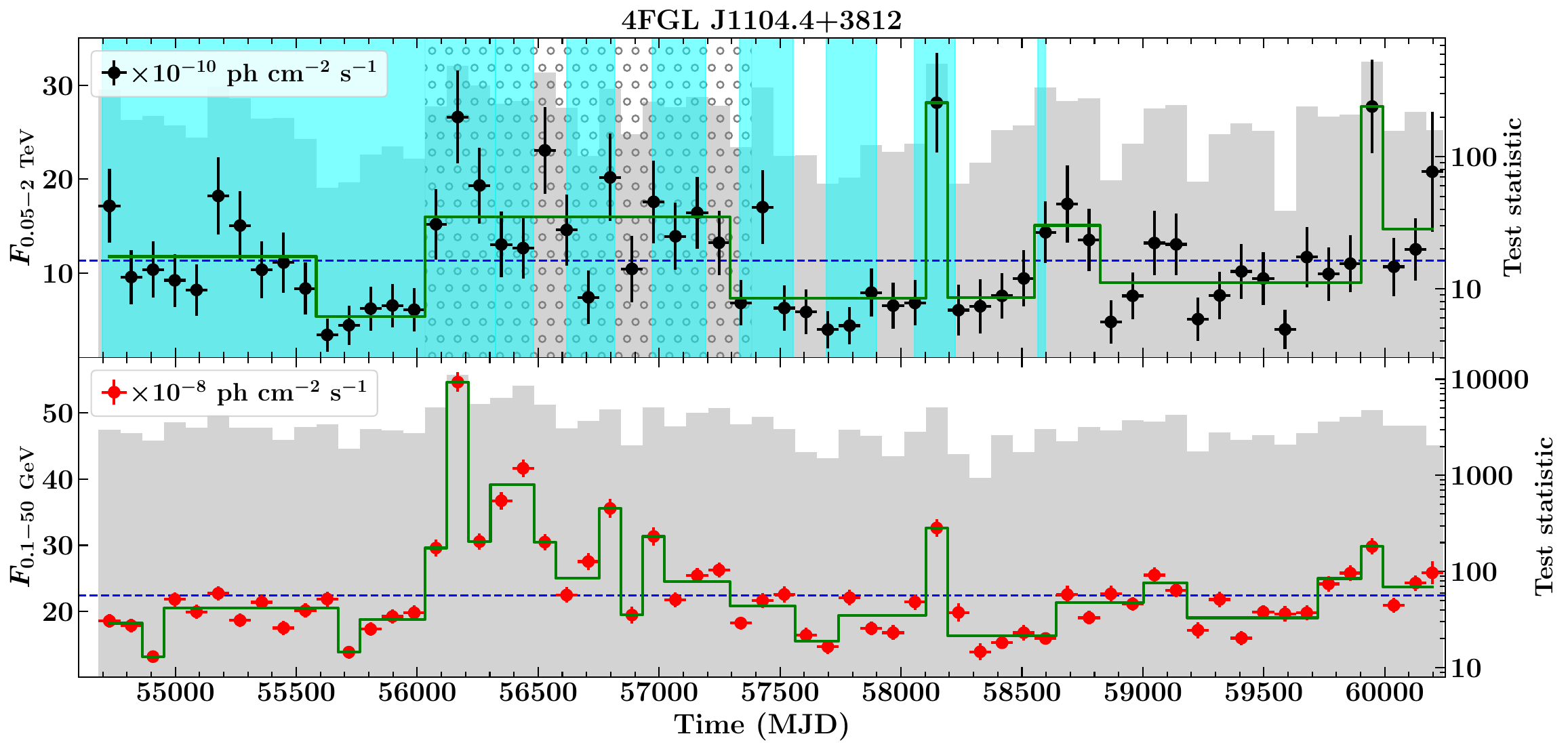}
\includegraphics[scale=0.23]{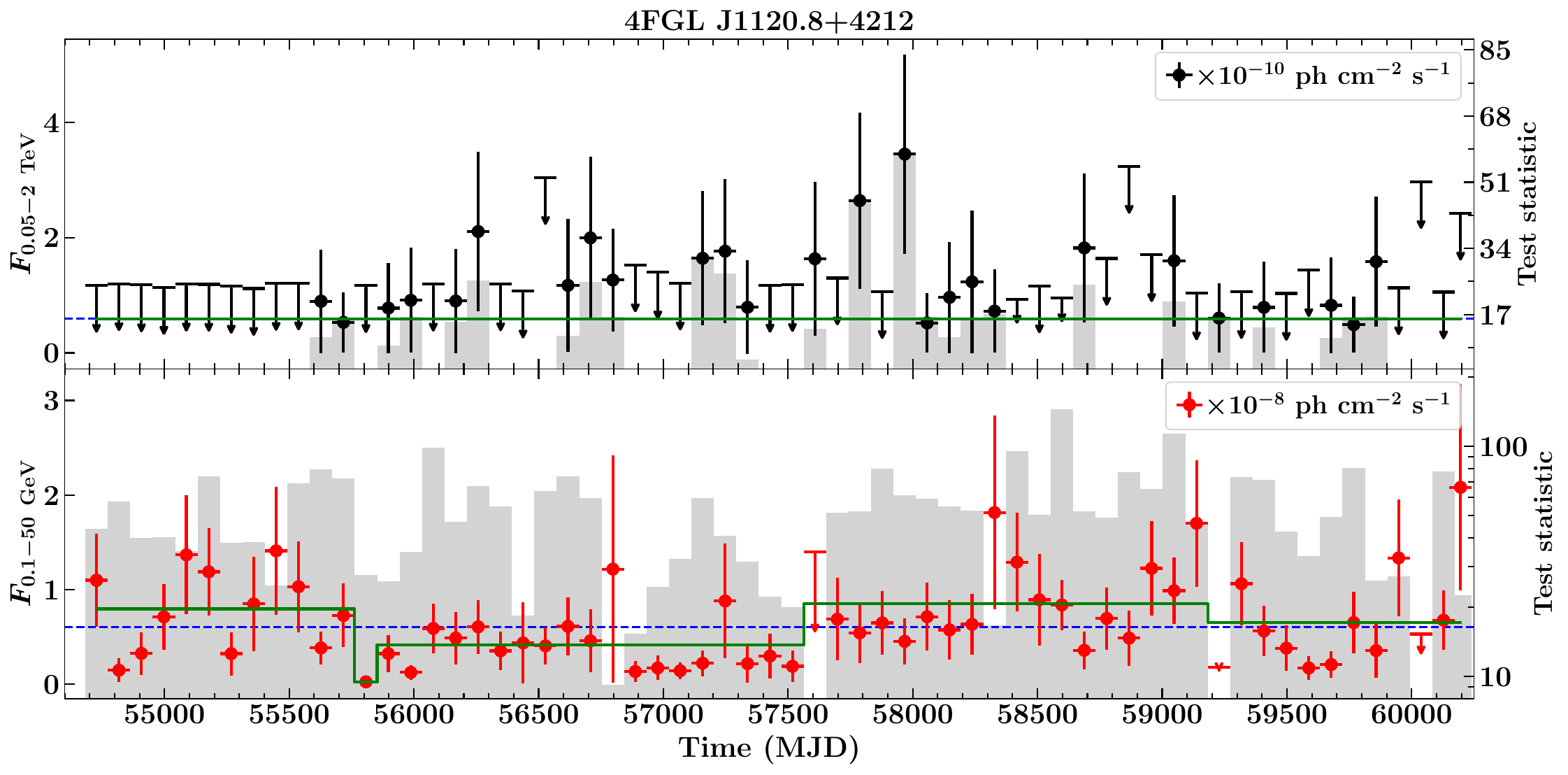}
}
\hbox{
\includegraphics[scale=0.23]{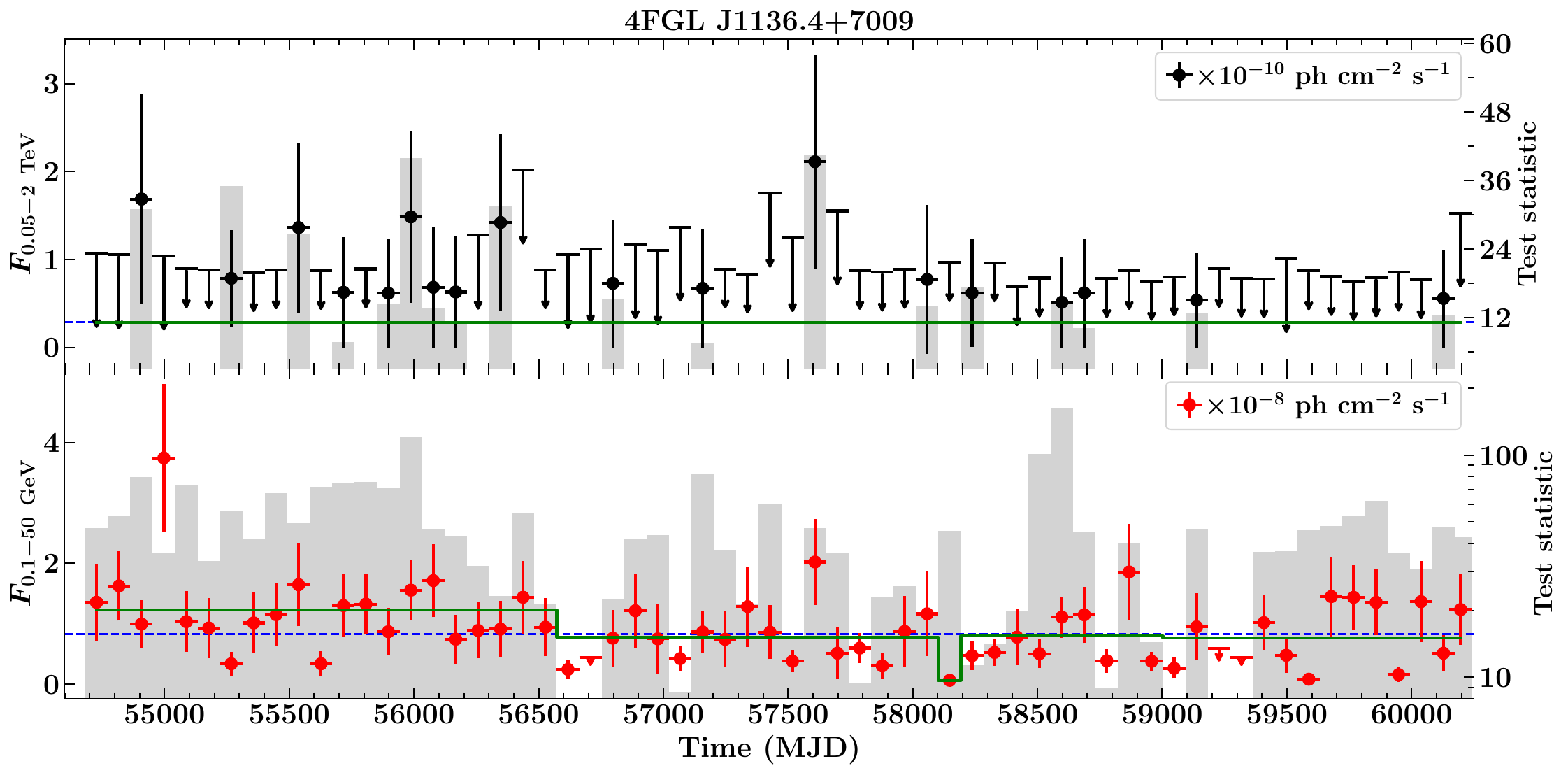}
\includegraphics[scale=0.23]{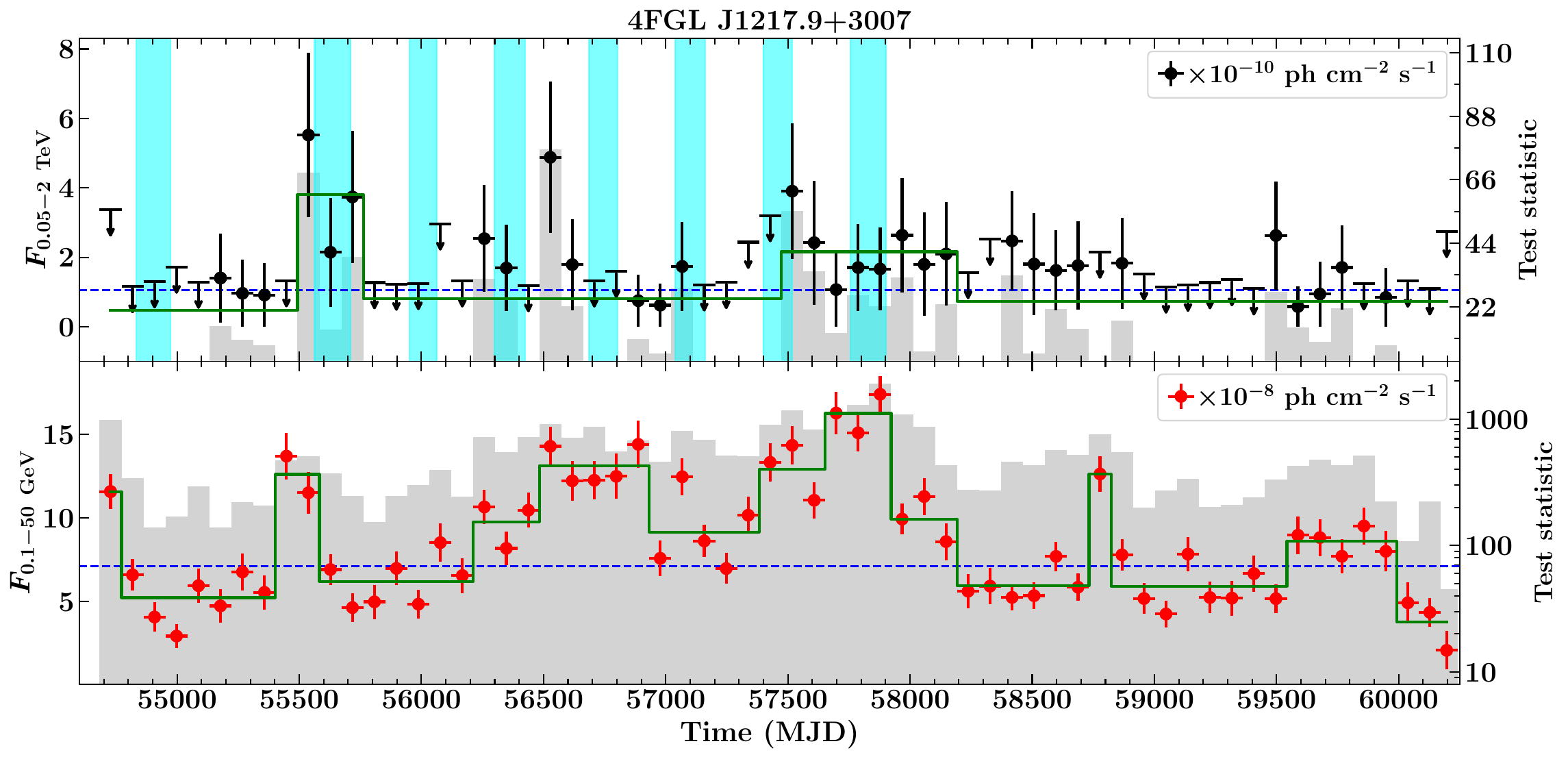}
}
\hbox{
\includegraphics[scale=0.23]{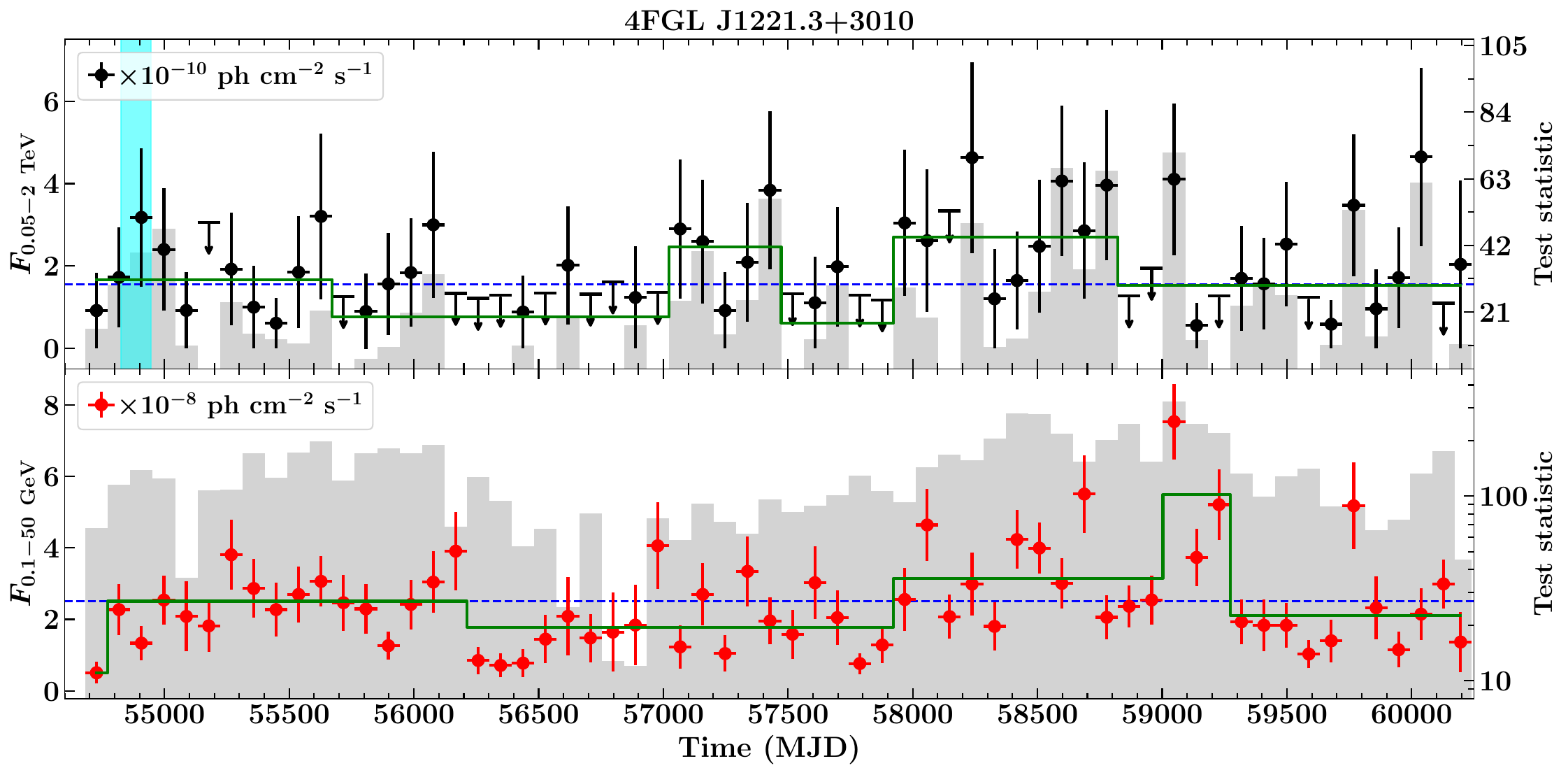}
\includegraphics[scale=0.23]{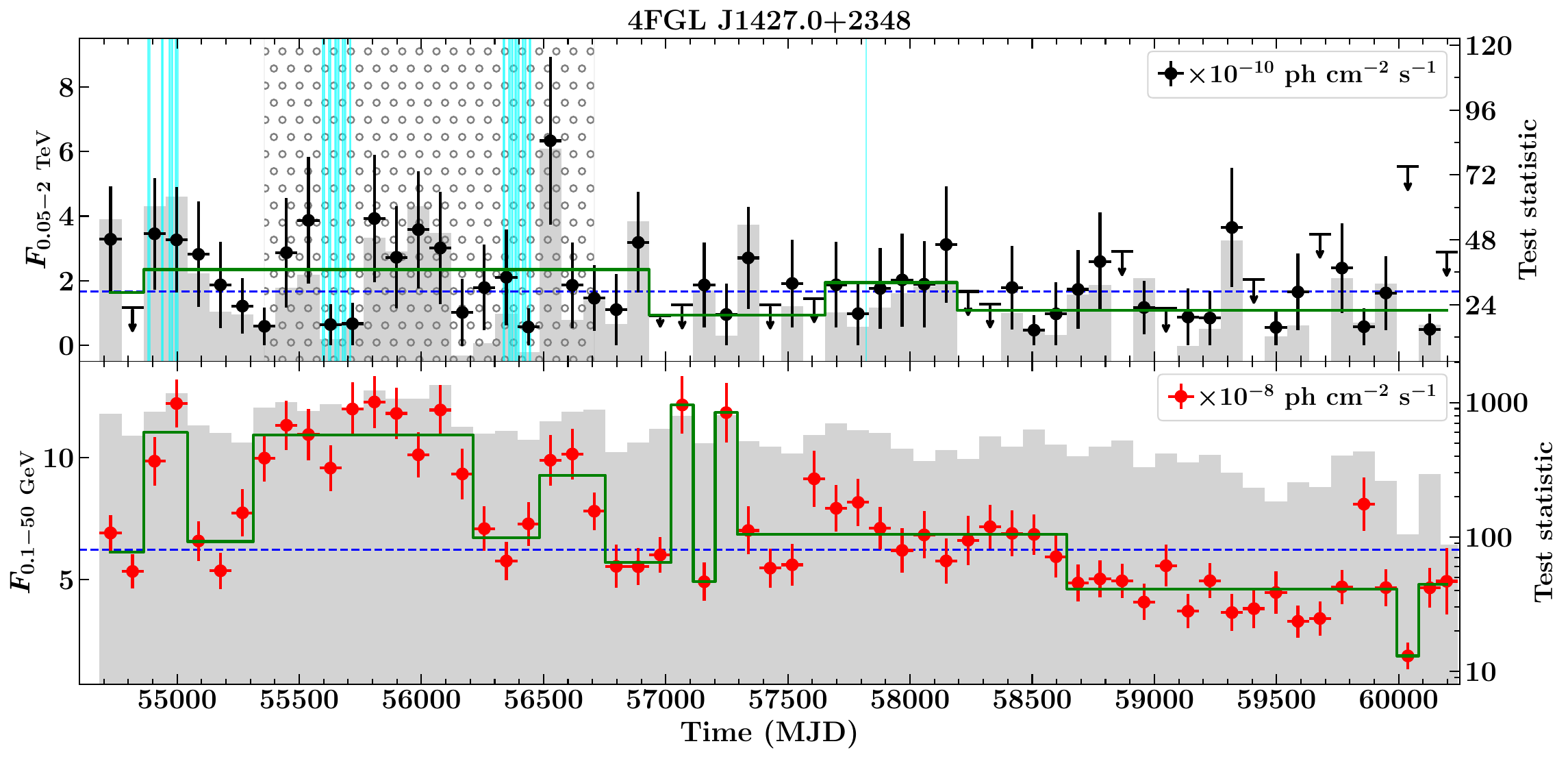}
}
\hbox{
\includegraphics[scale=0.23]{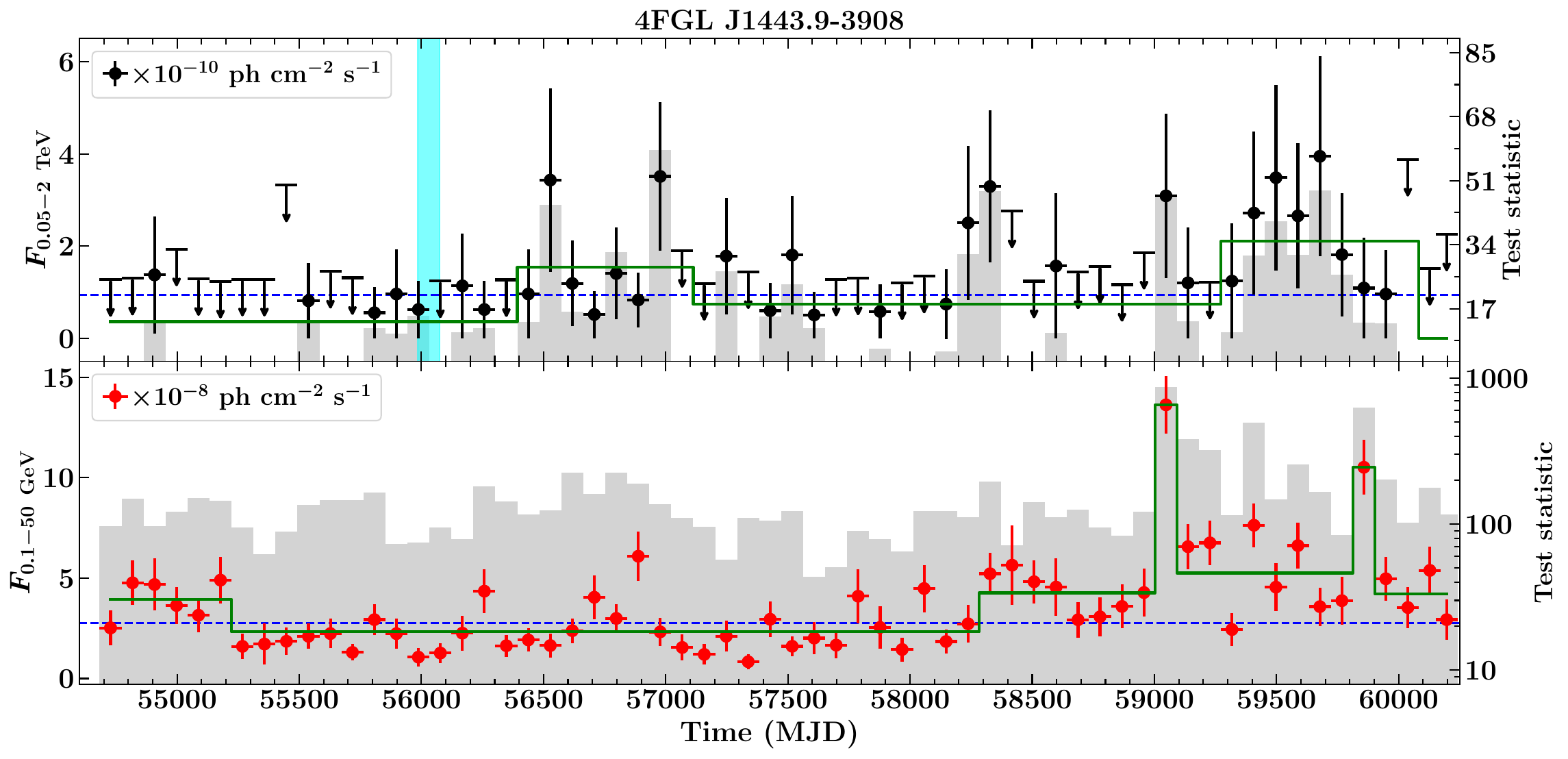}
\includegraphics[scale=0.23]{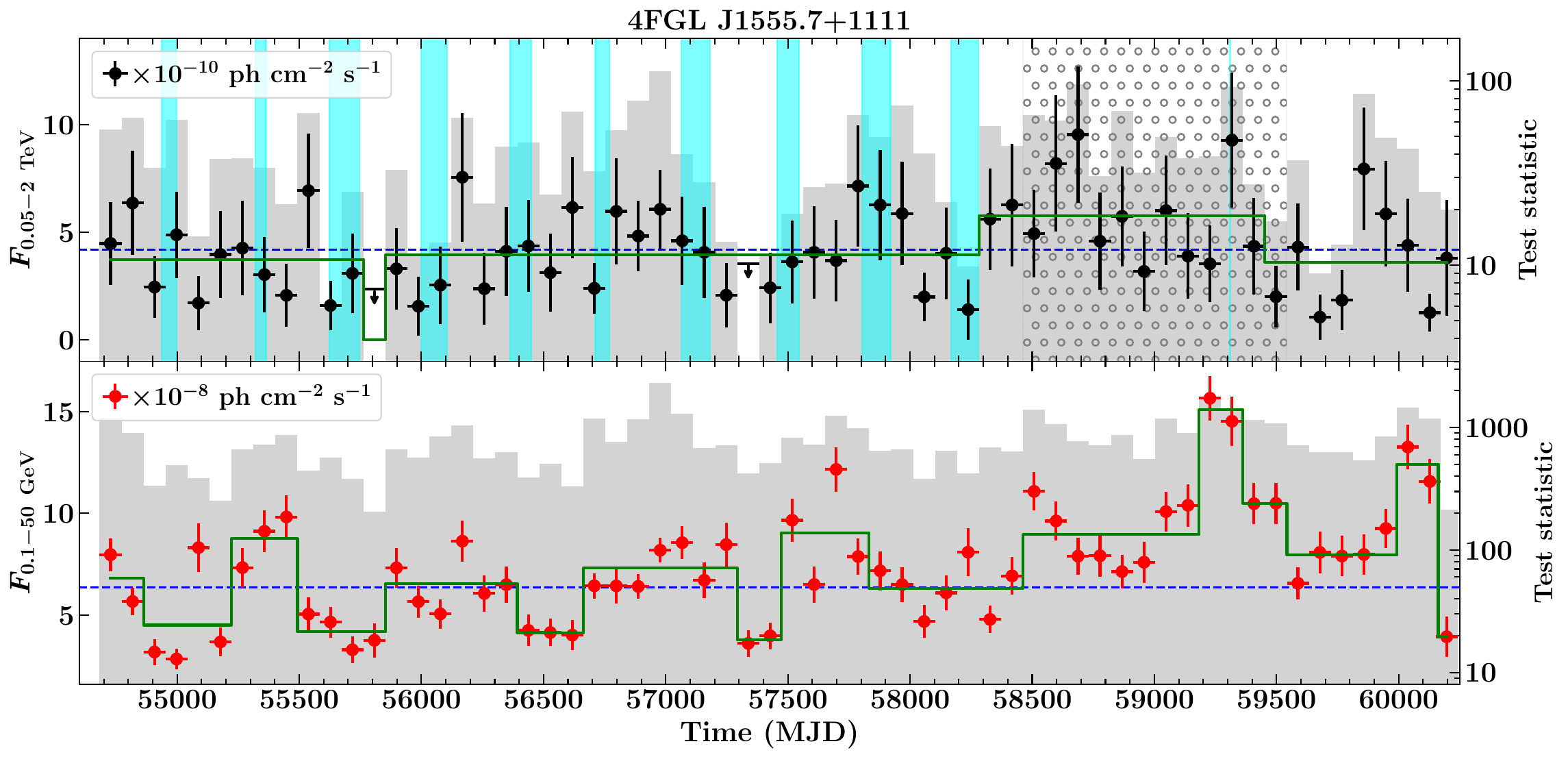}
}

\caption{Figure 1 (continued)}
\end{figure*}

\begin{figure*}
\figurenum{1}
\hbox{
\includegraphics[scale=0.23]{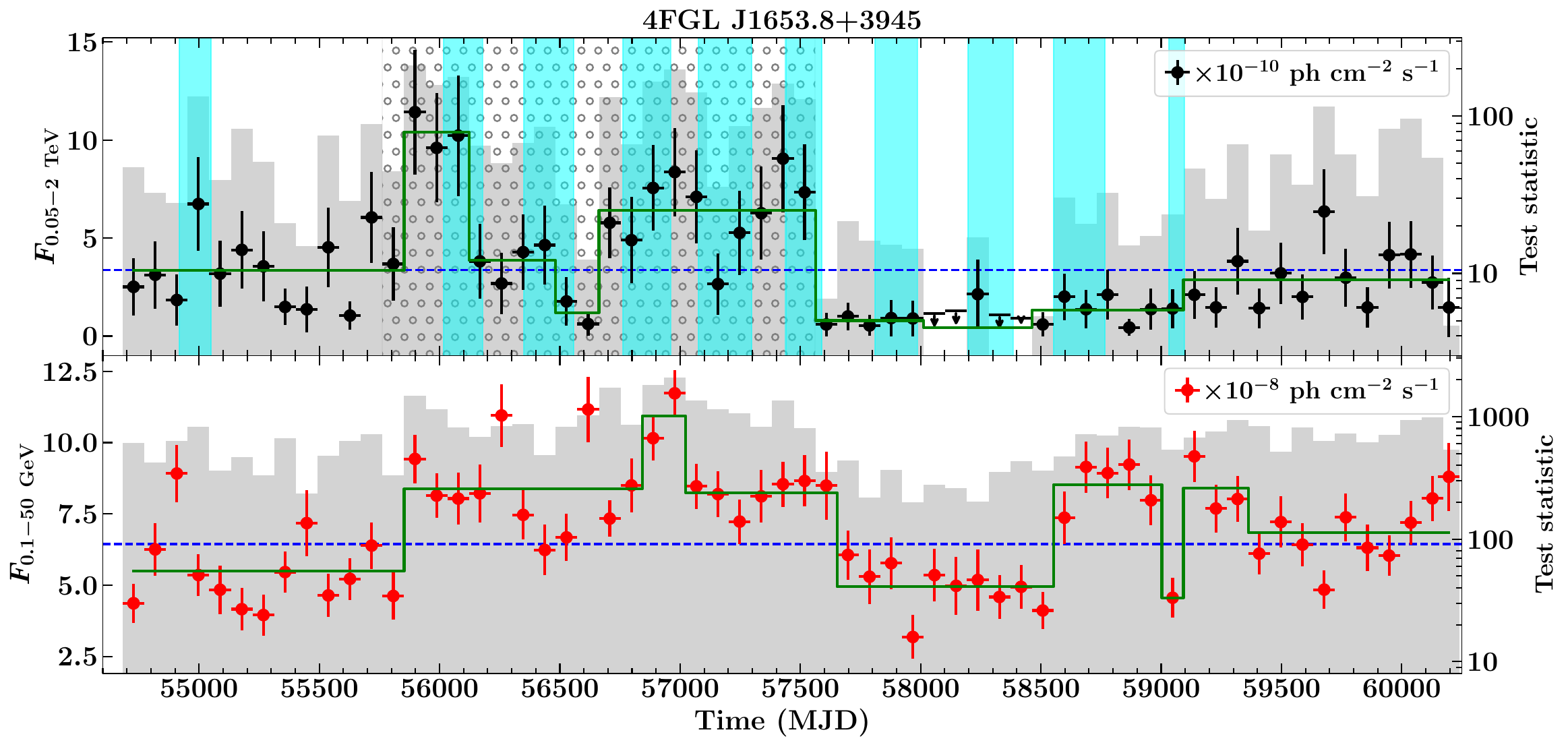}
\includegraphics[scale=0.23]{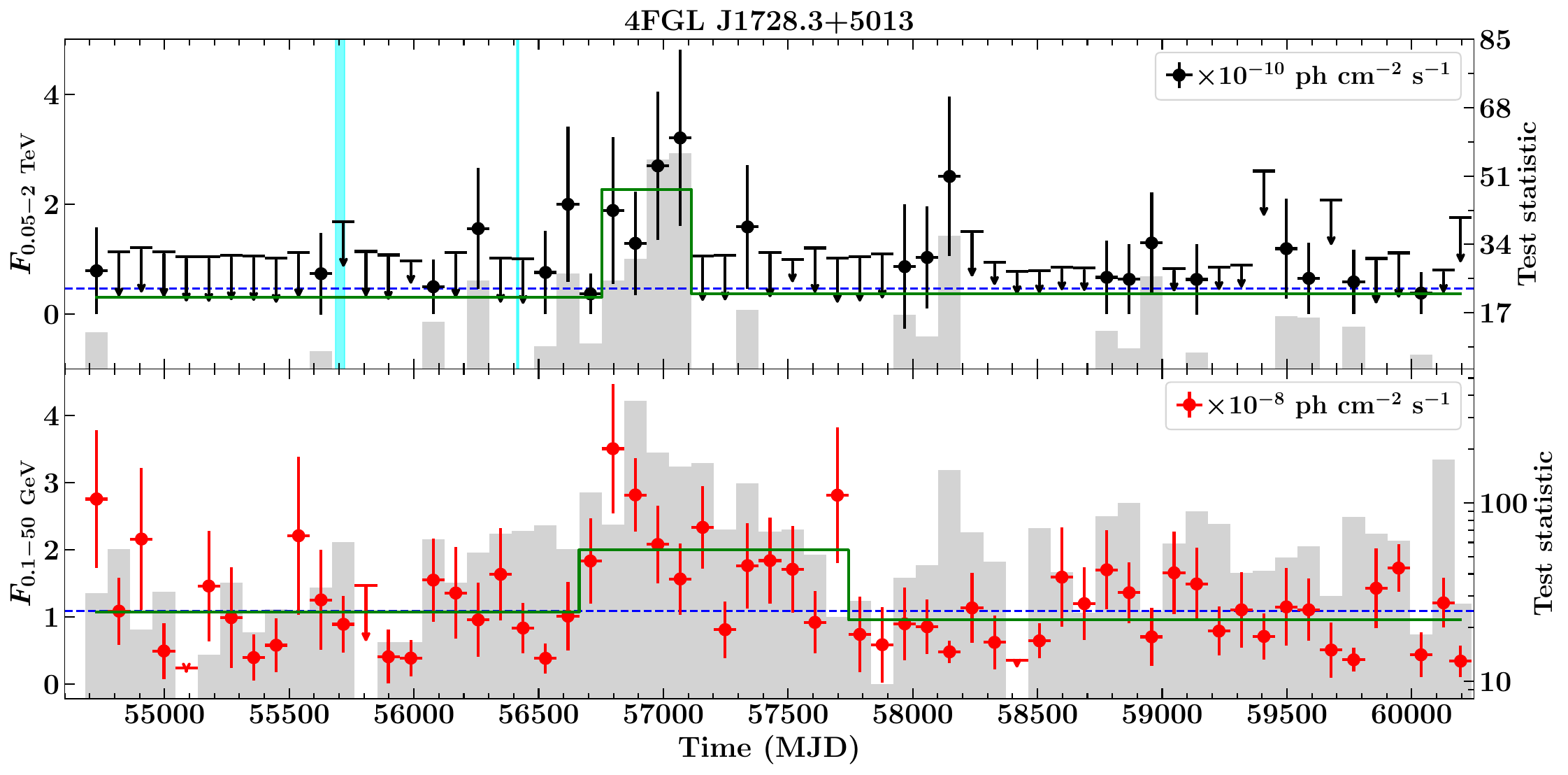}
}
\hbox{
\includegraphics[scale=0.23]{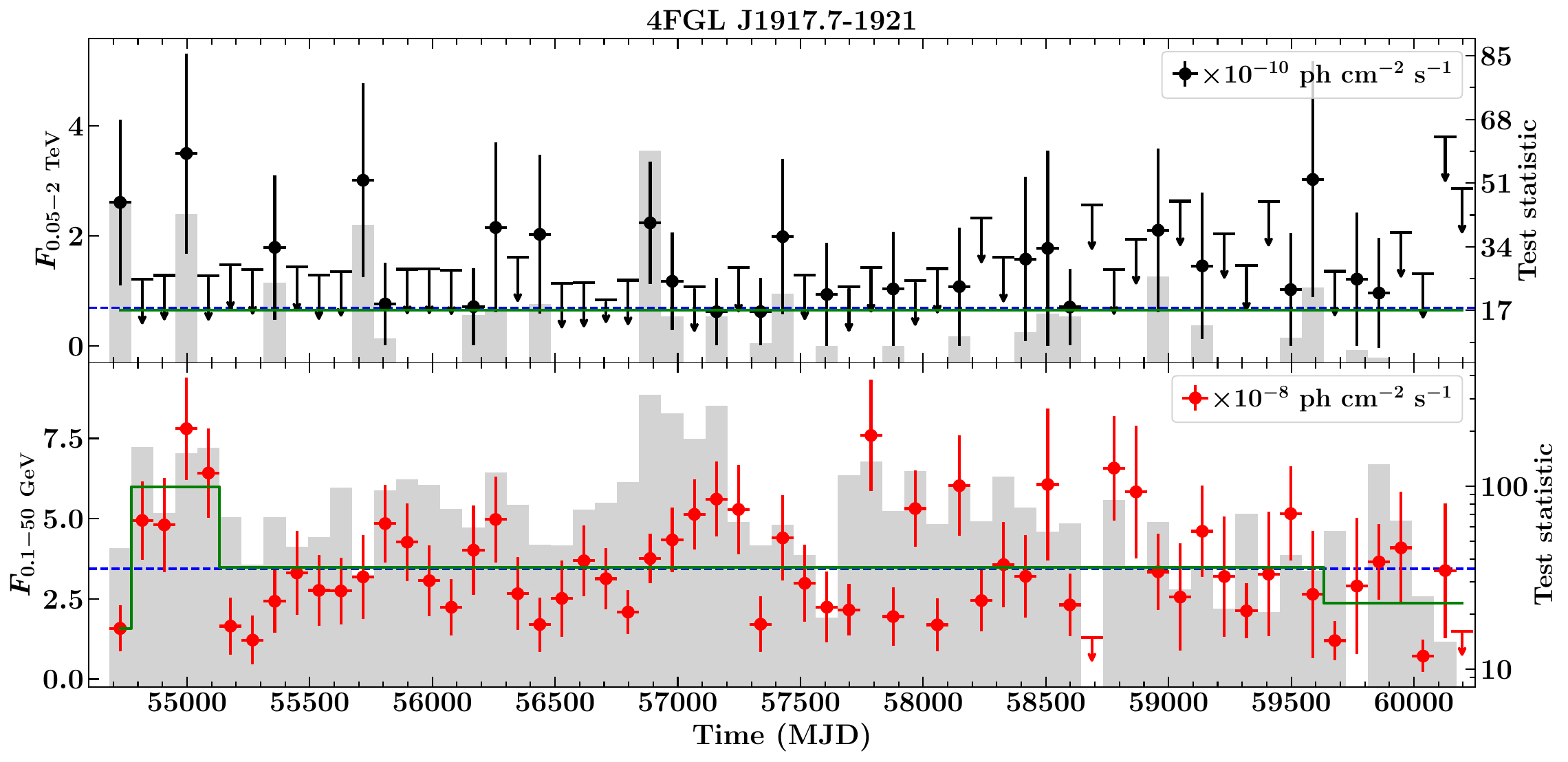}
\includegraphics[scale=0.23]{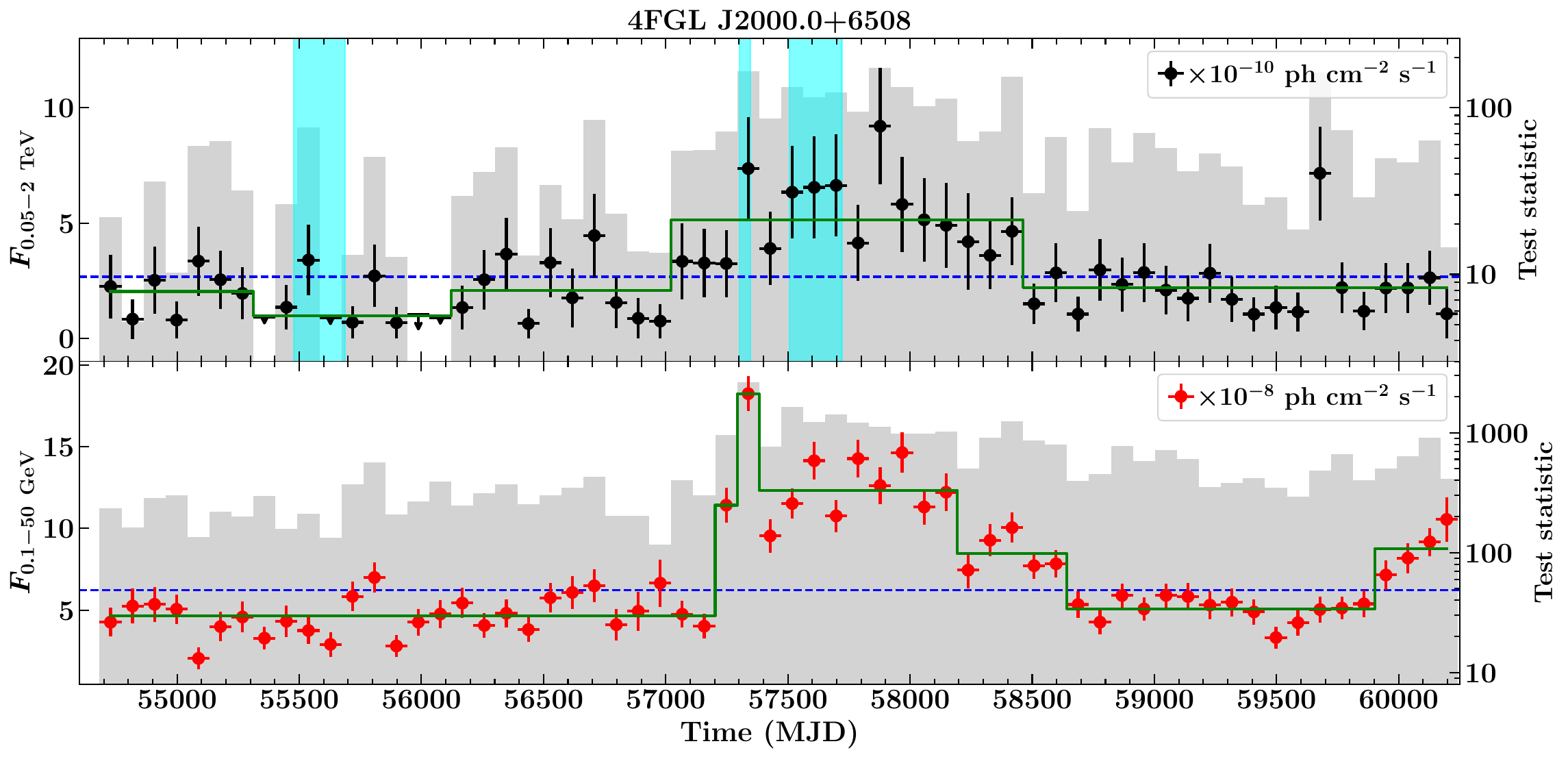}
}
\hbox{
\includegraphics[scale=0.23]{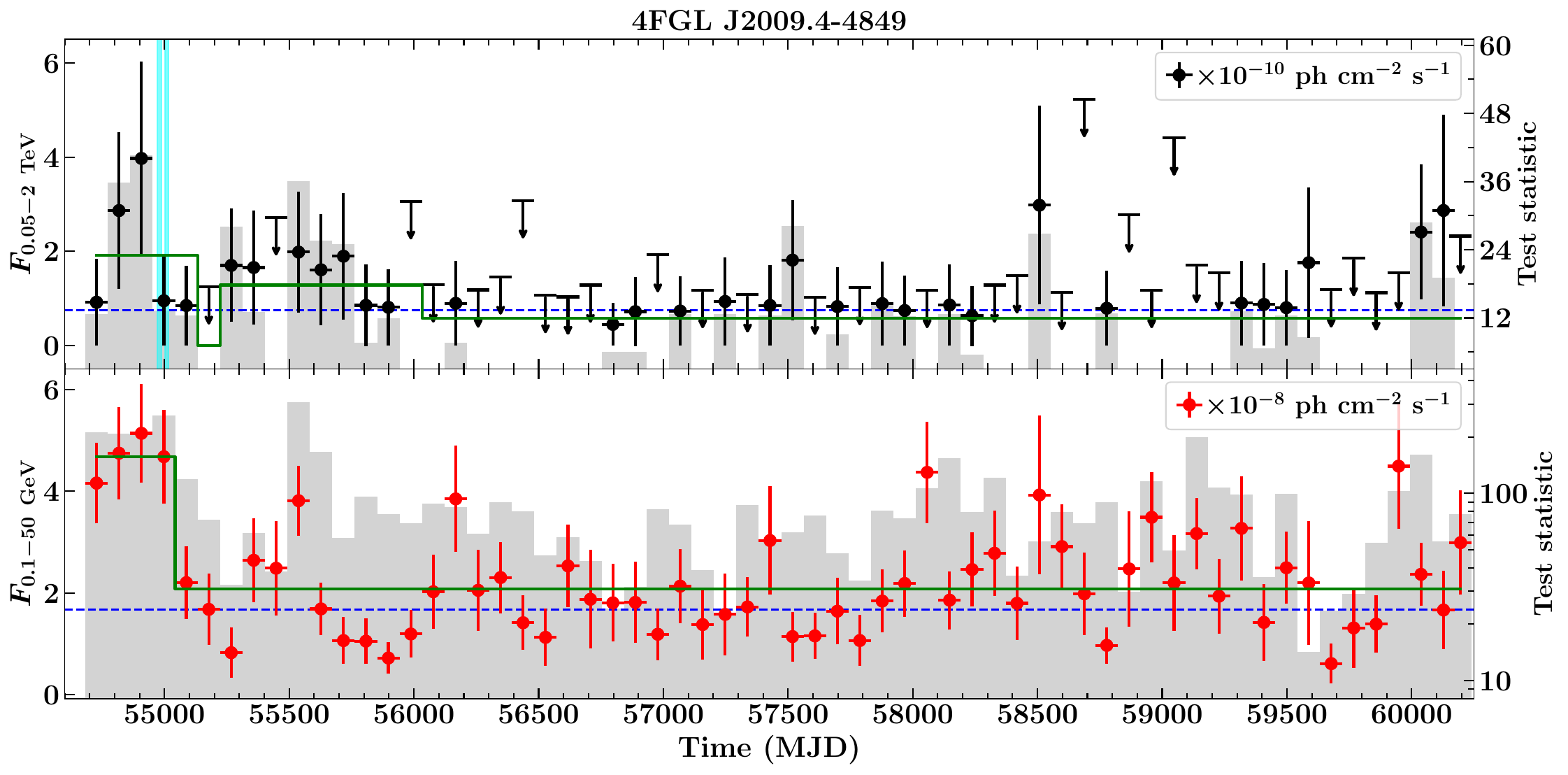}
\includegraphics[scale=0.23]{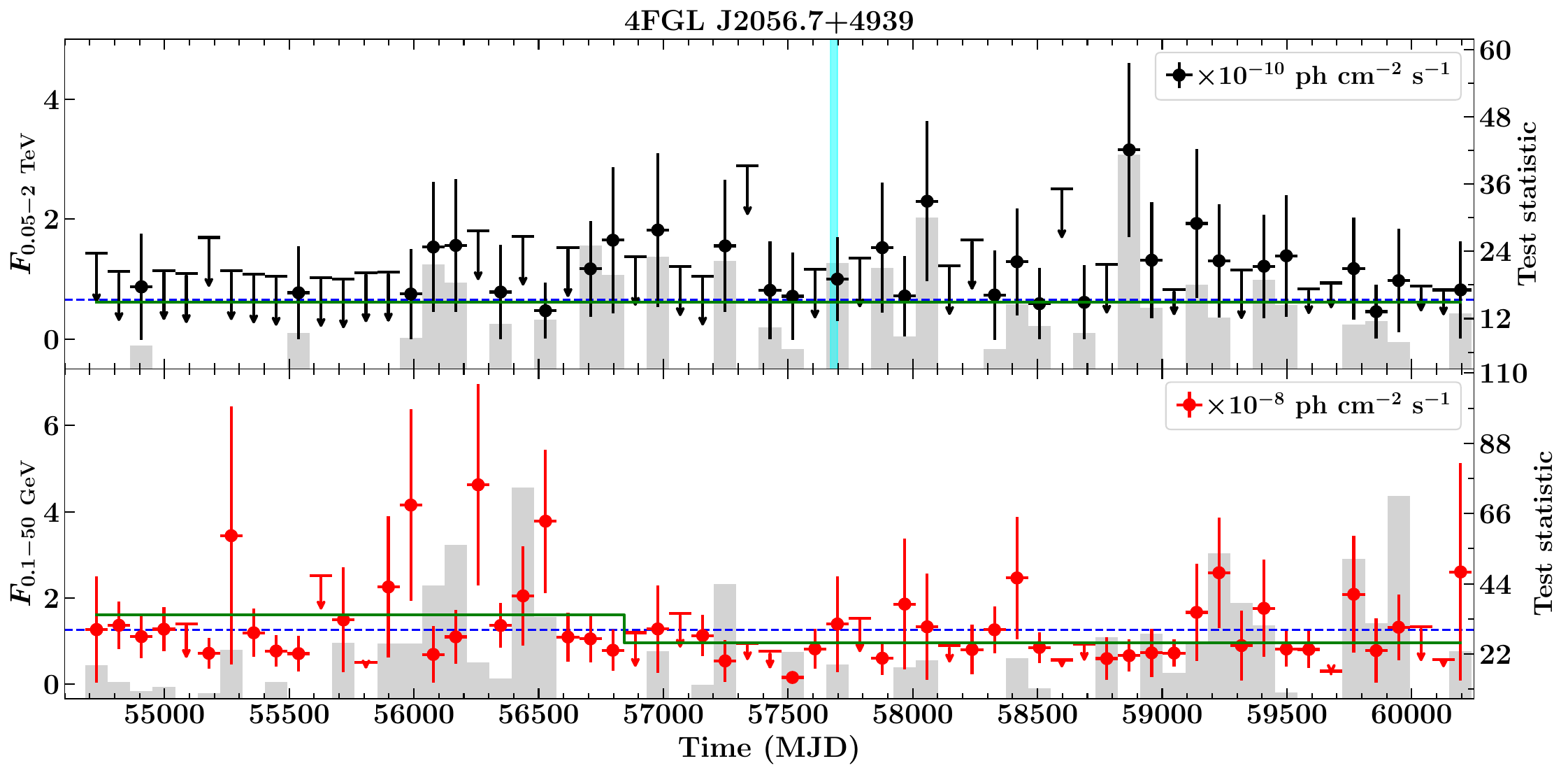}
}
\hbox{
\includegraphics[scale=0.23]{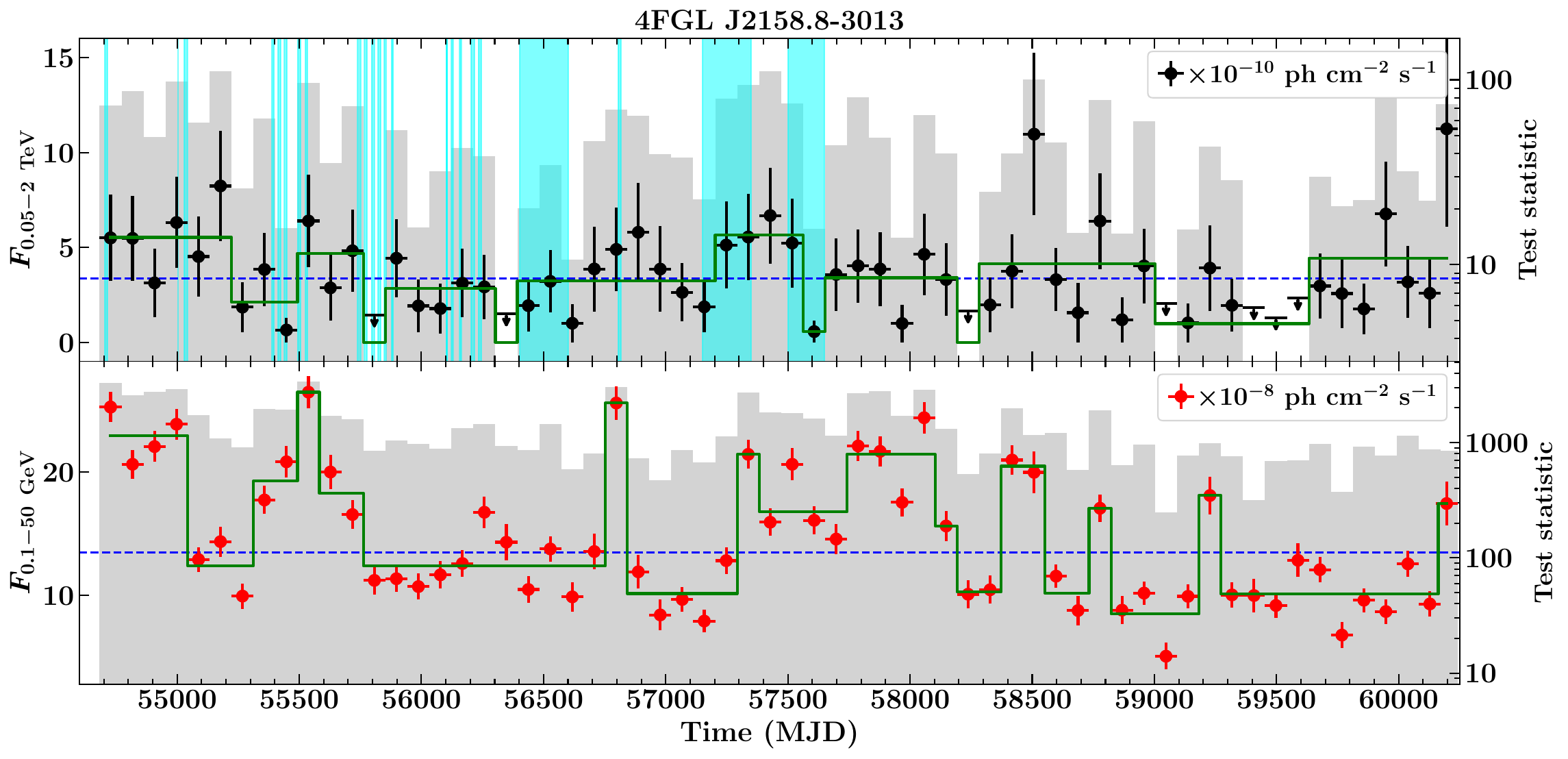}
\includegraphics[scale=0.23]{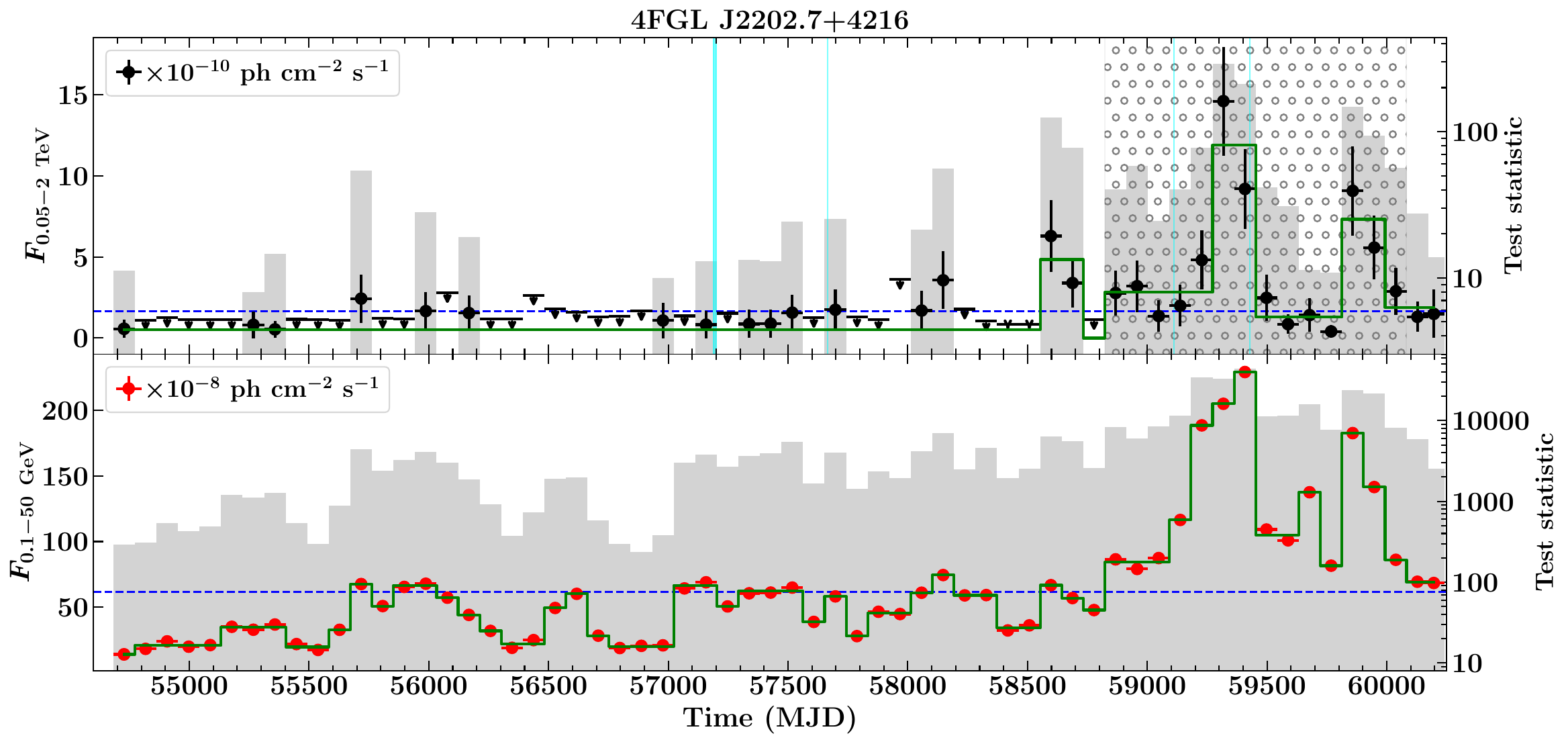}
}
\hbox{
\includegraphics[scale=0.23]{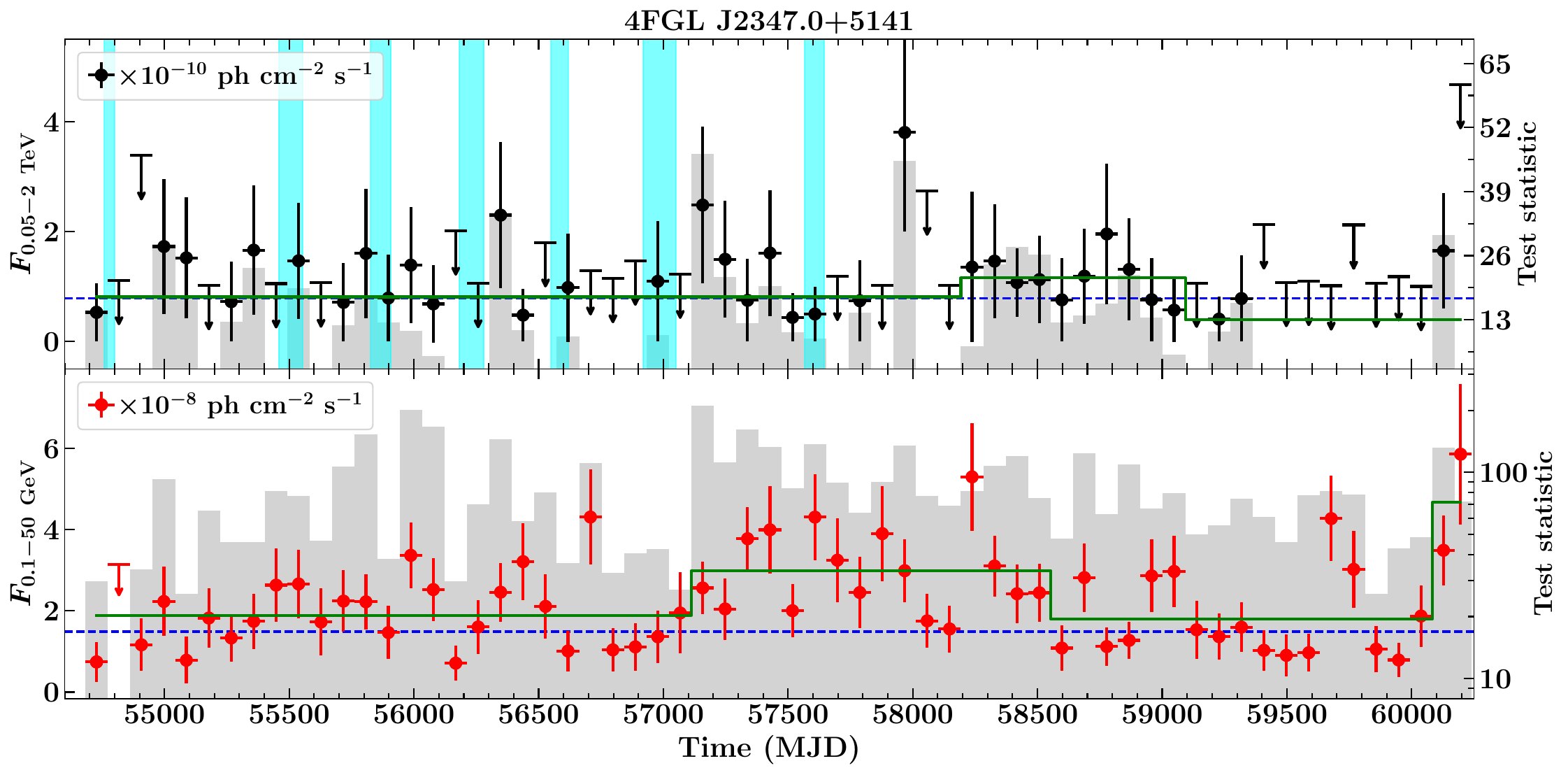}
}

\caption{Figure 1 (continued)}
\end{figure*}

\begin{figure*}
\hbox{
\includegraphics[scale=0.23]{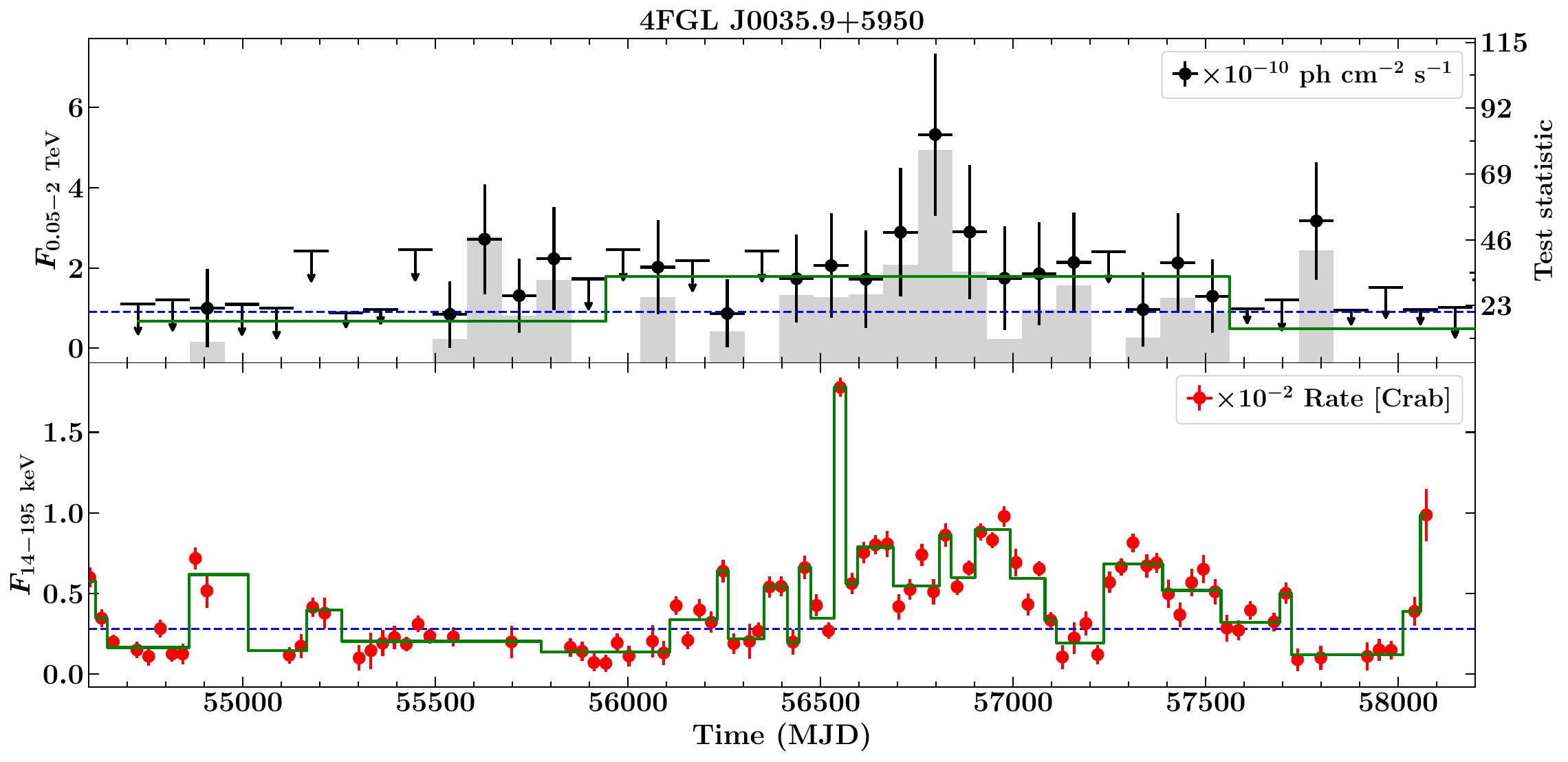}
\includegraphics[scale=0.23]{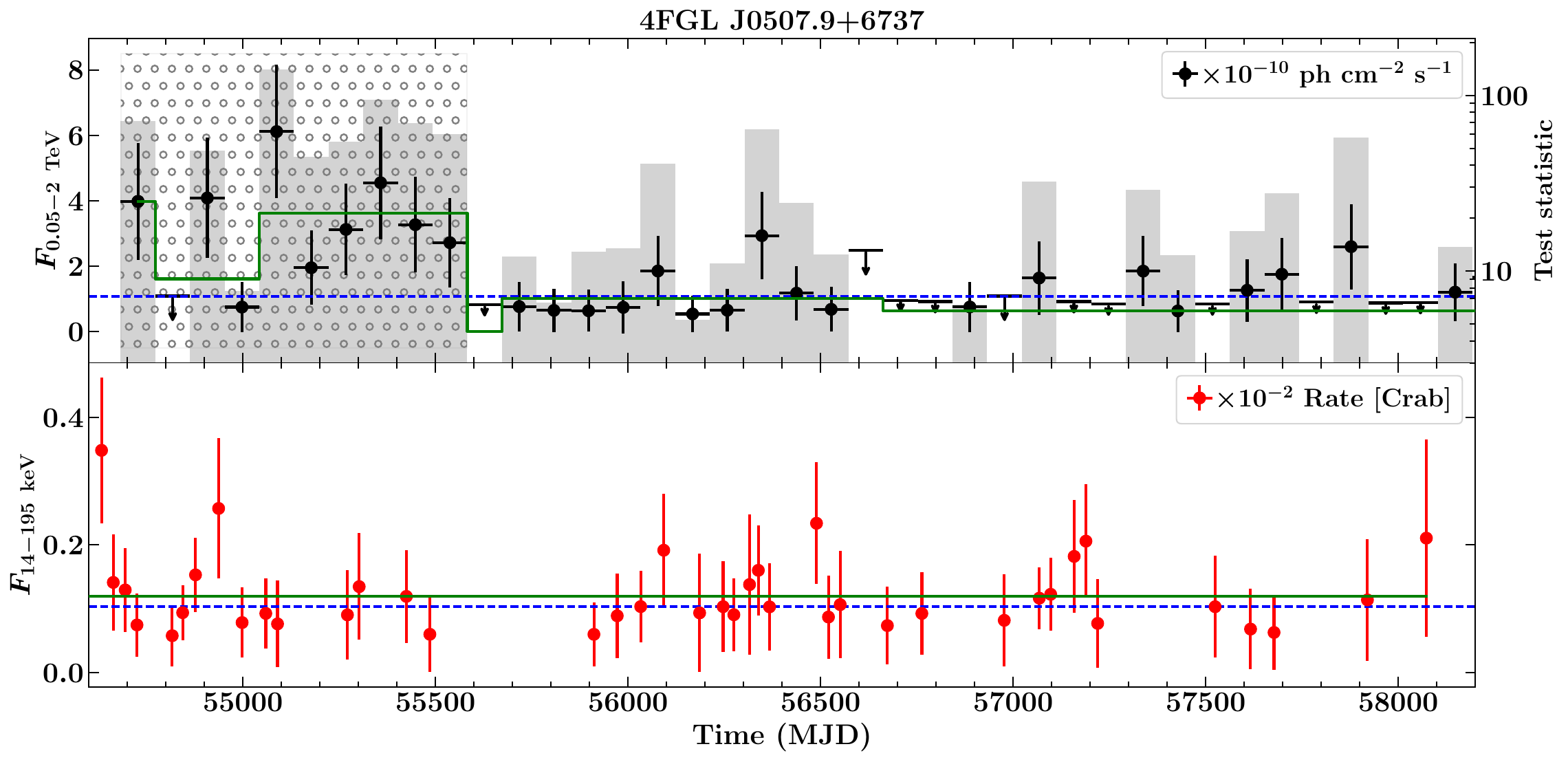}
}
\hbox{
\includegraphics[scale=0.23]{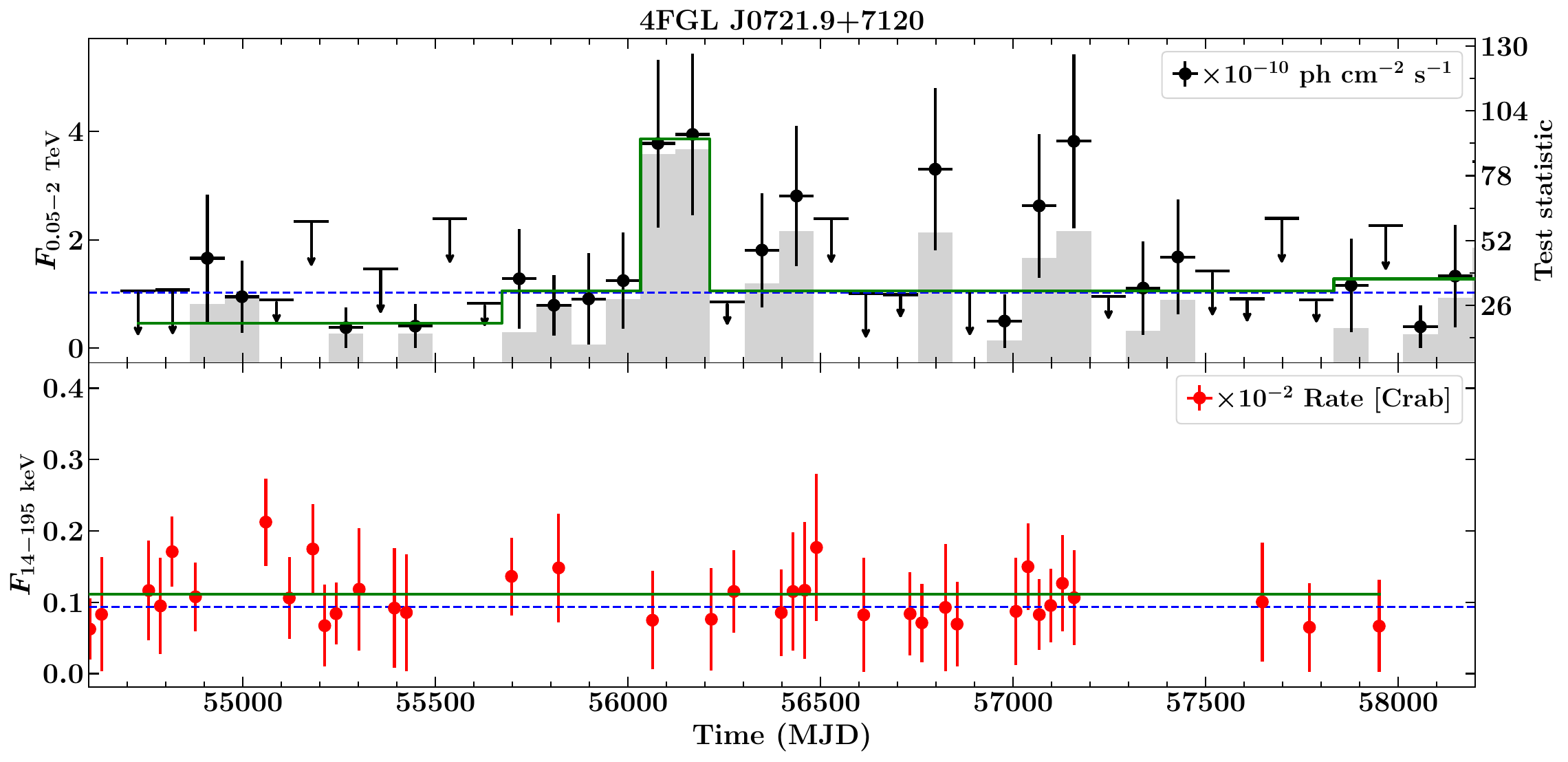}
\includegraphics[scale=0.23]{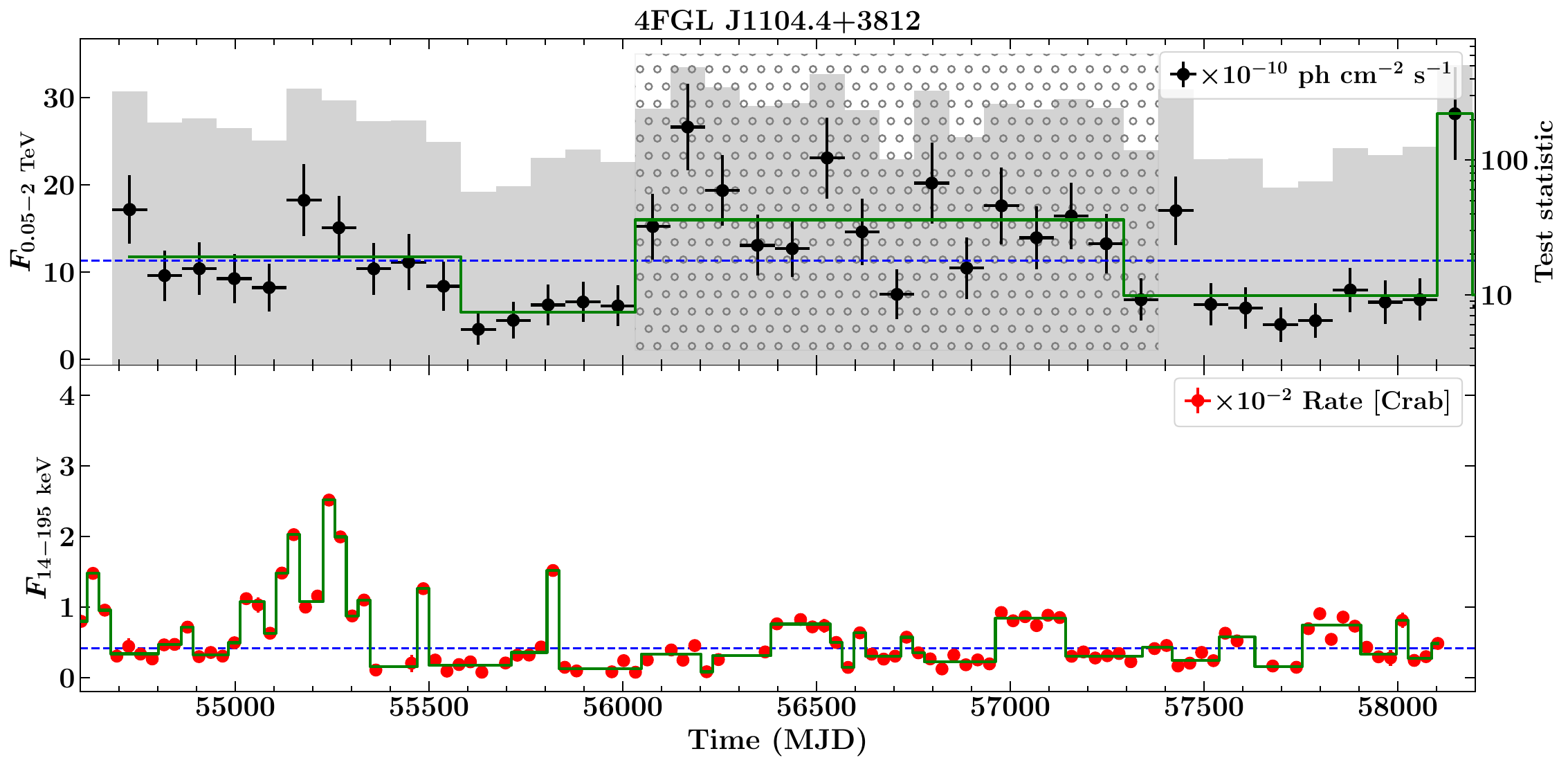}
}
\hbox{
\includegraphics[scale=0.23]{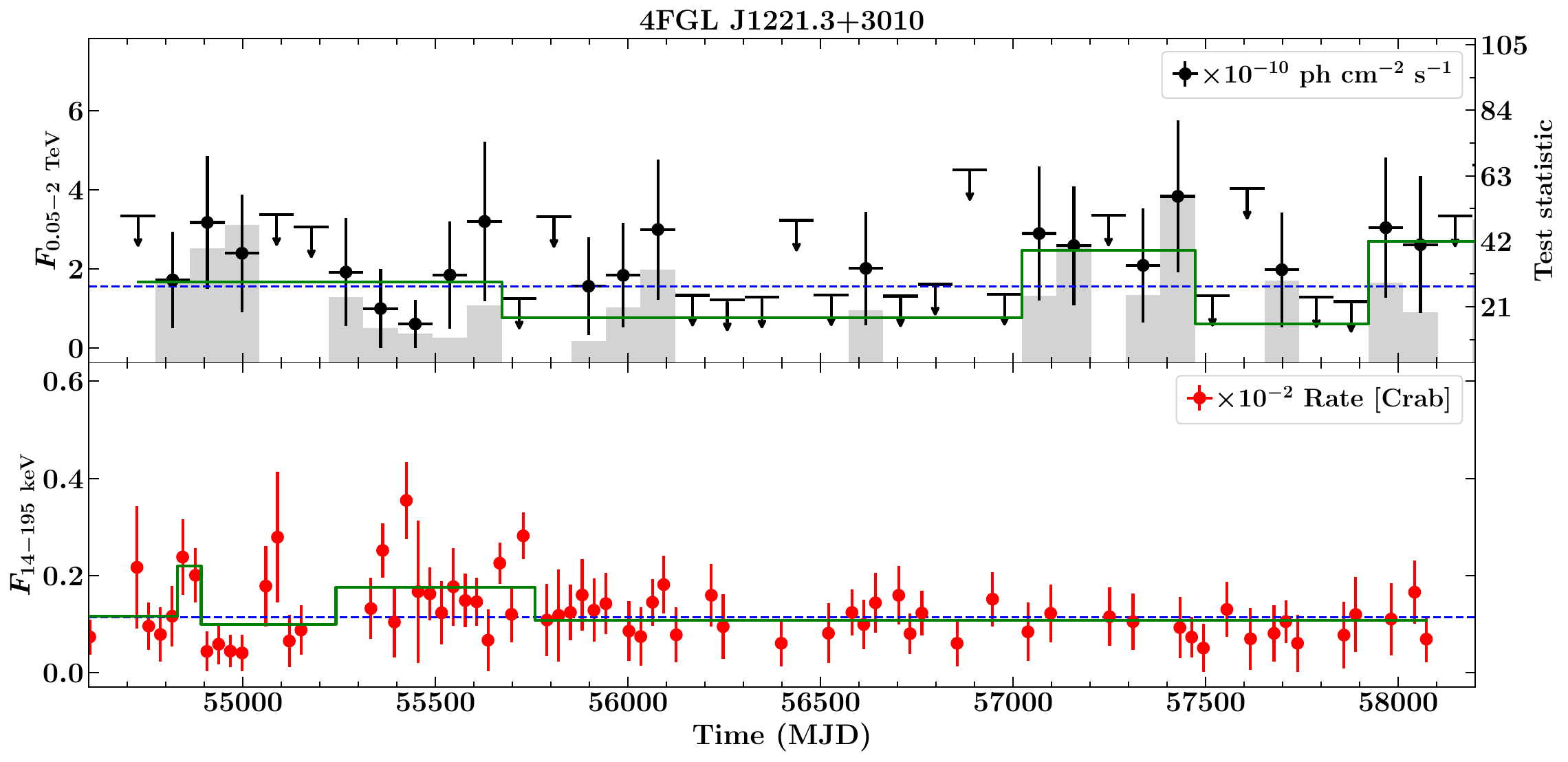}
\includegraphics[scale=0.23]{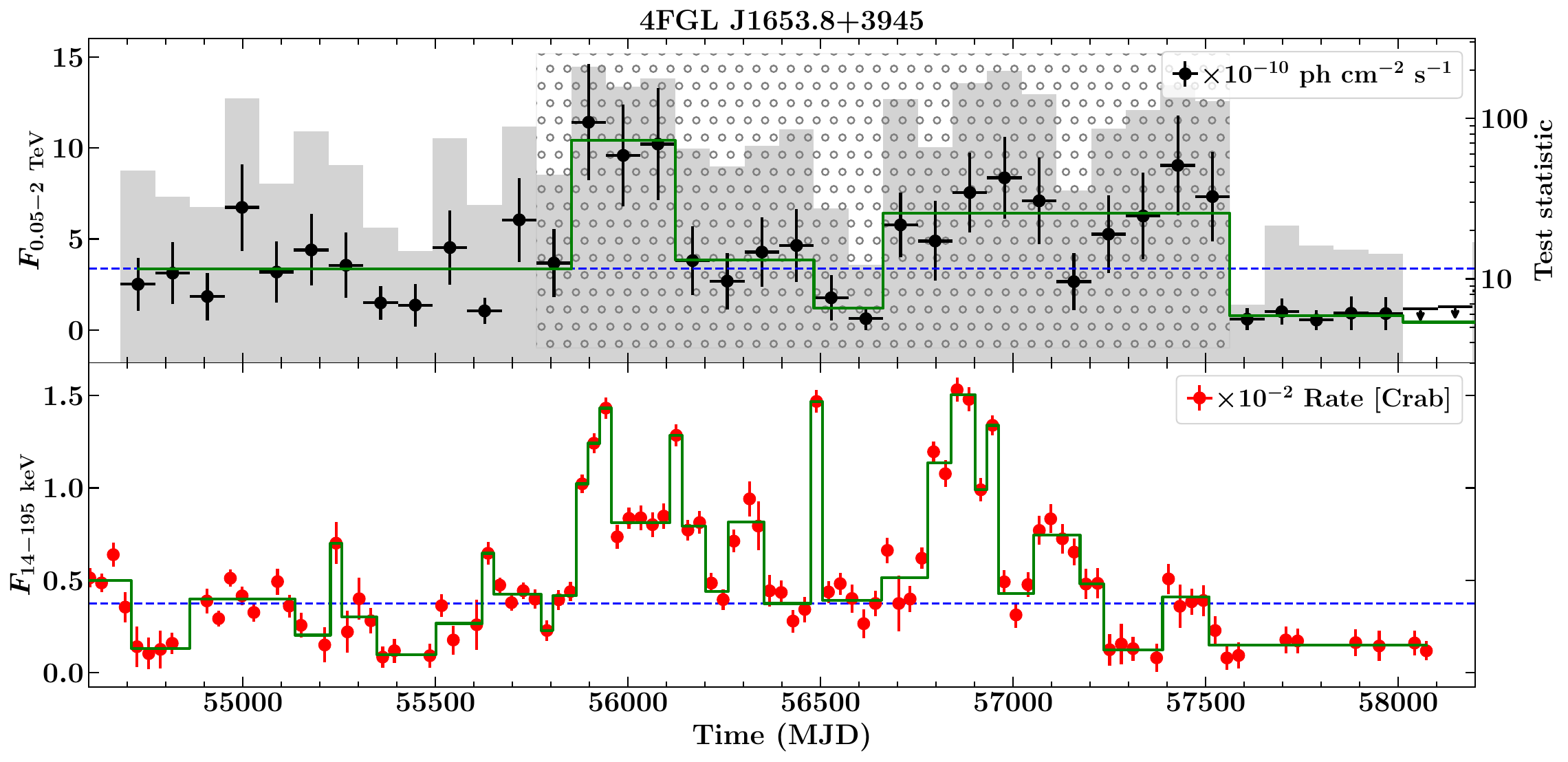}
}
\hbox{
\includegraphics[scale=0.23]{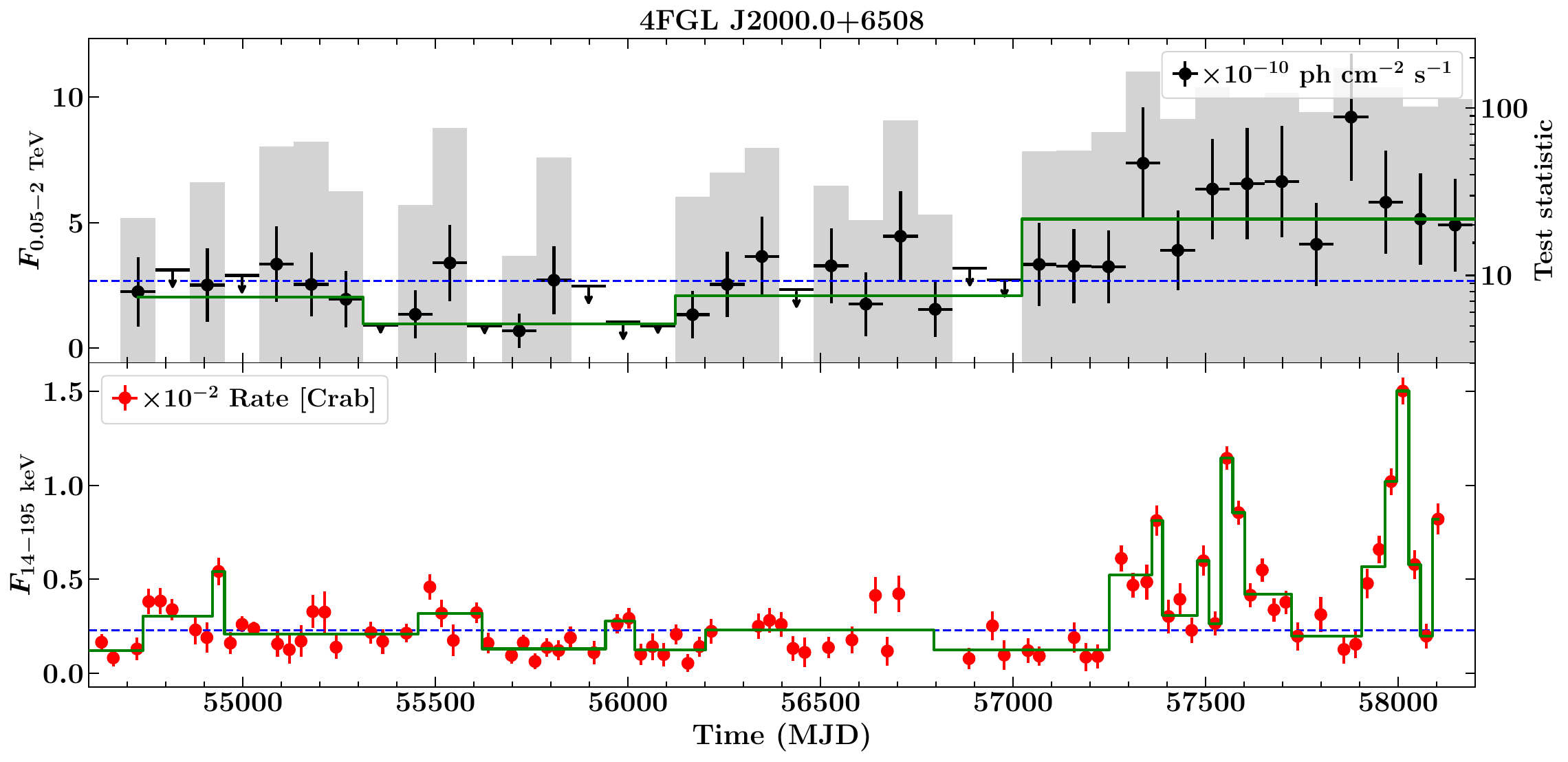}
\includegraphics[scale=0.23]{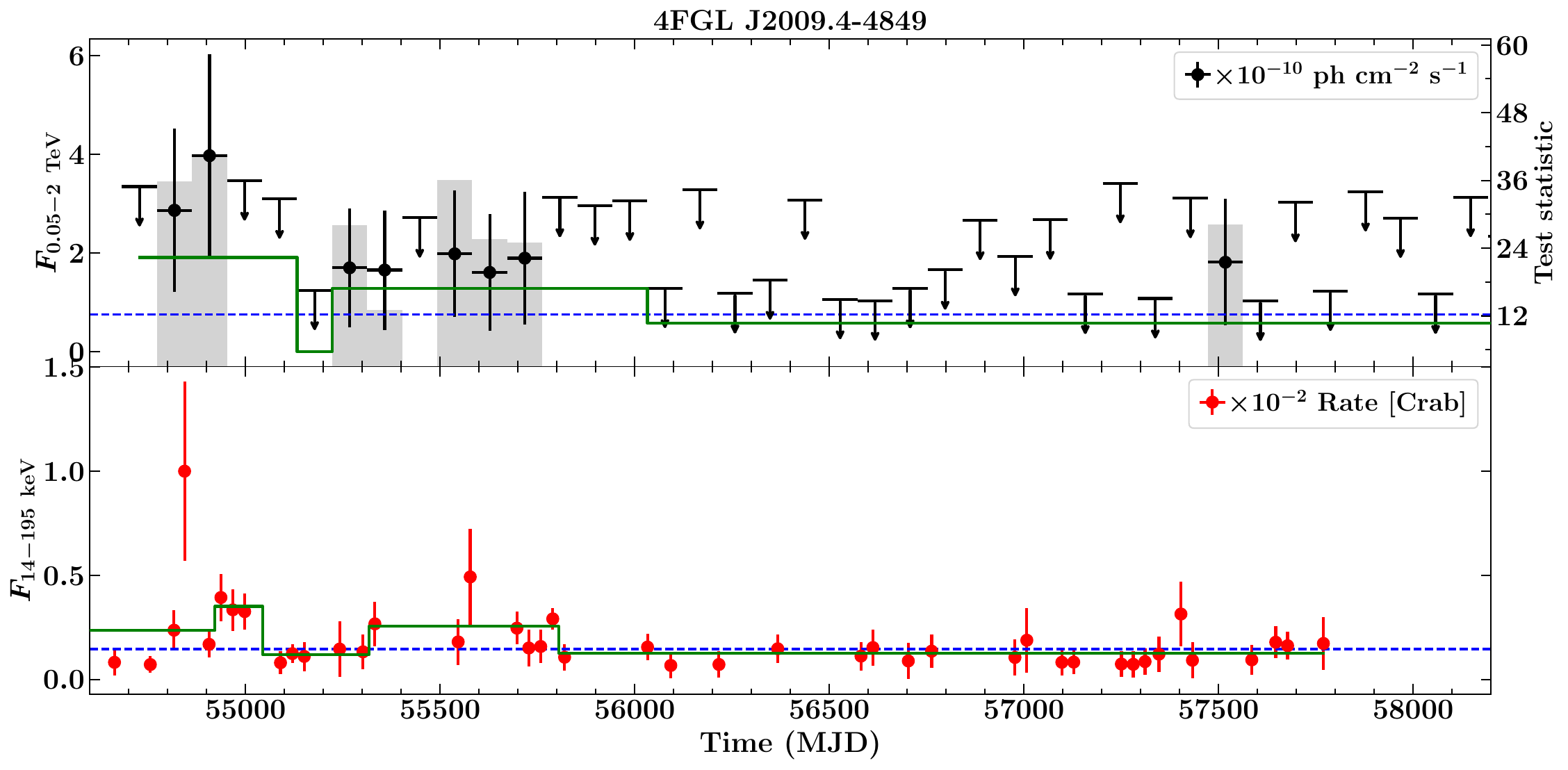}
}
\hbox{
\includegraphics[scale=0.23]{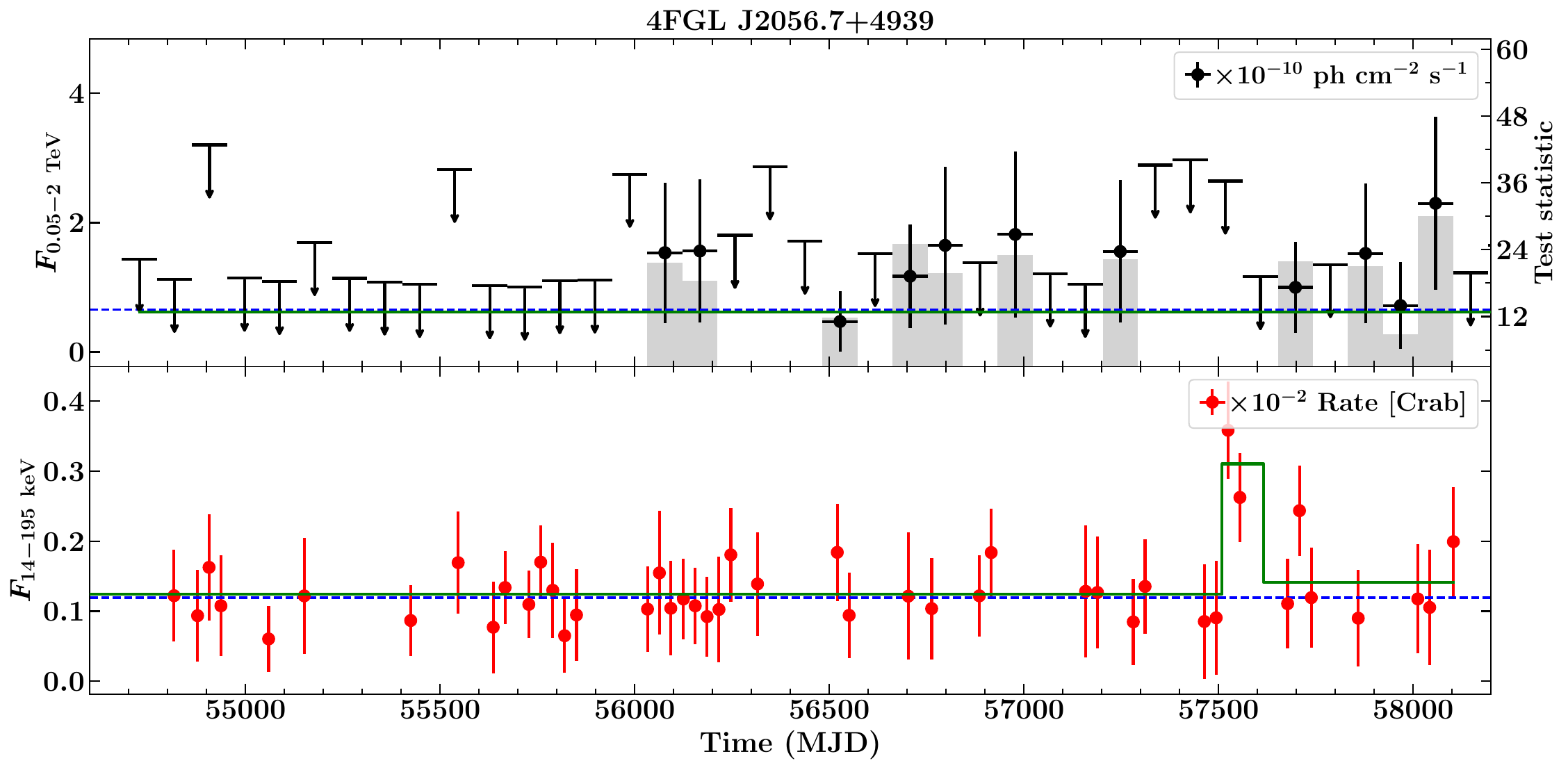}
\includegraphics[scale=0.23]{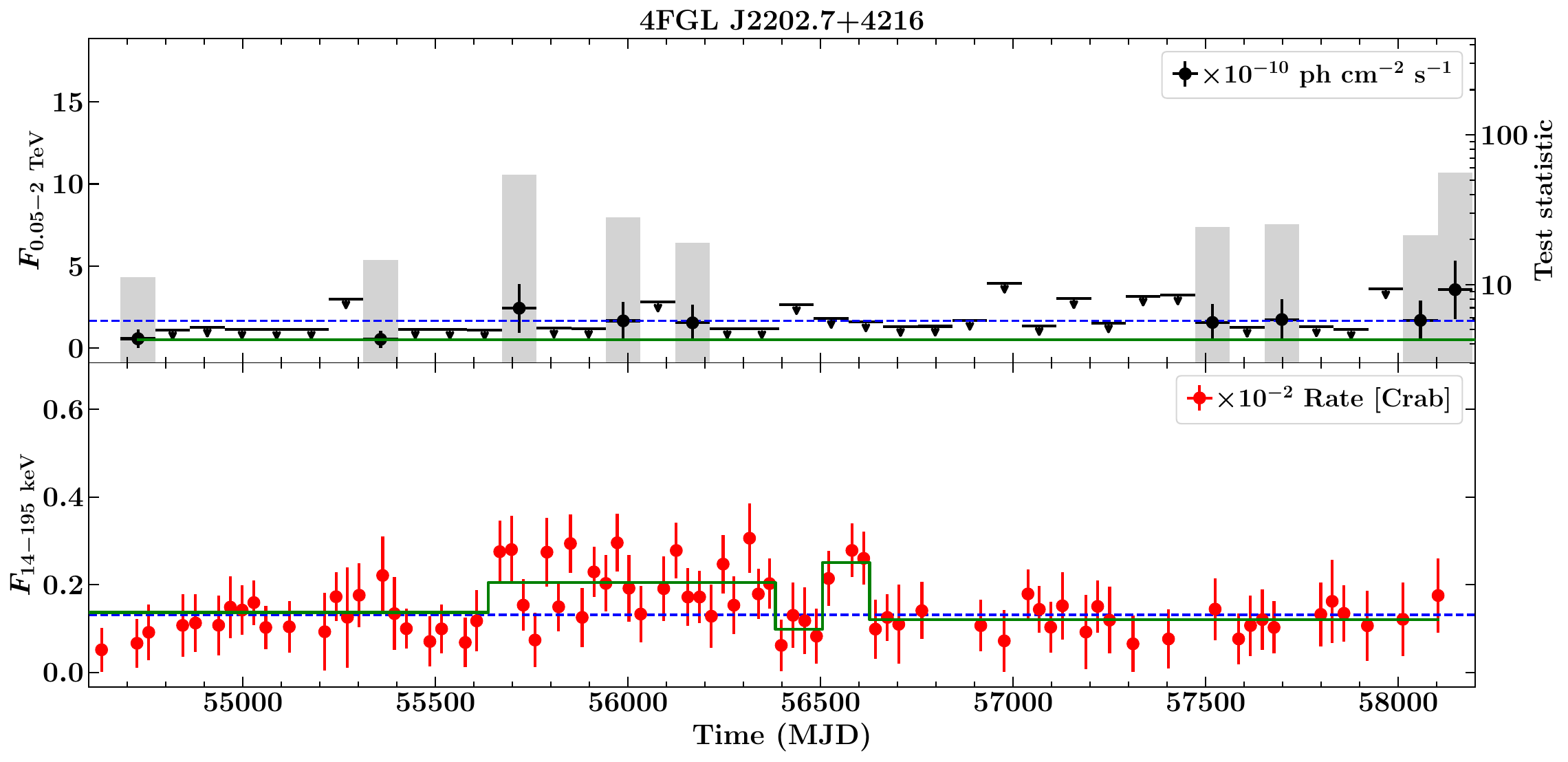}
}

\caption{A comparison of the TeV and hard X-ray flux variations.  The bottom panel shows the Crab weighted, monthly-binned light curve in 14$-$195 keV energy range.  Other information are same as in Figure~\ref{fig:1}.} \label{fig:2}
\end{figure*}

\begin{figure}
\figurenum{2}
\hbox{
\includegraphics[scale=0.23]{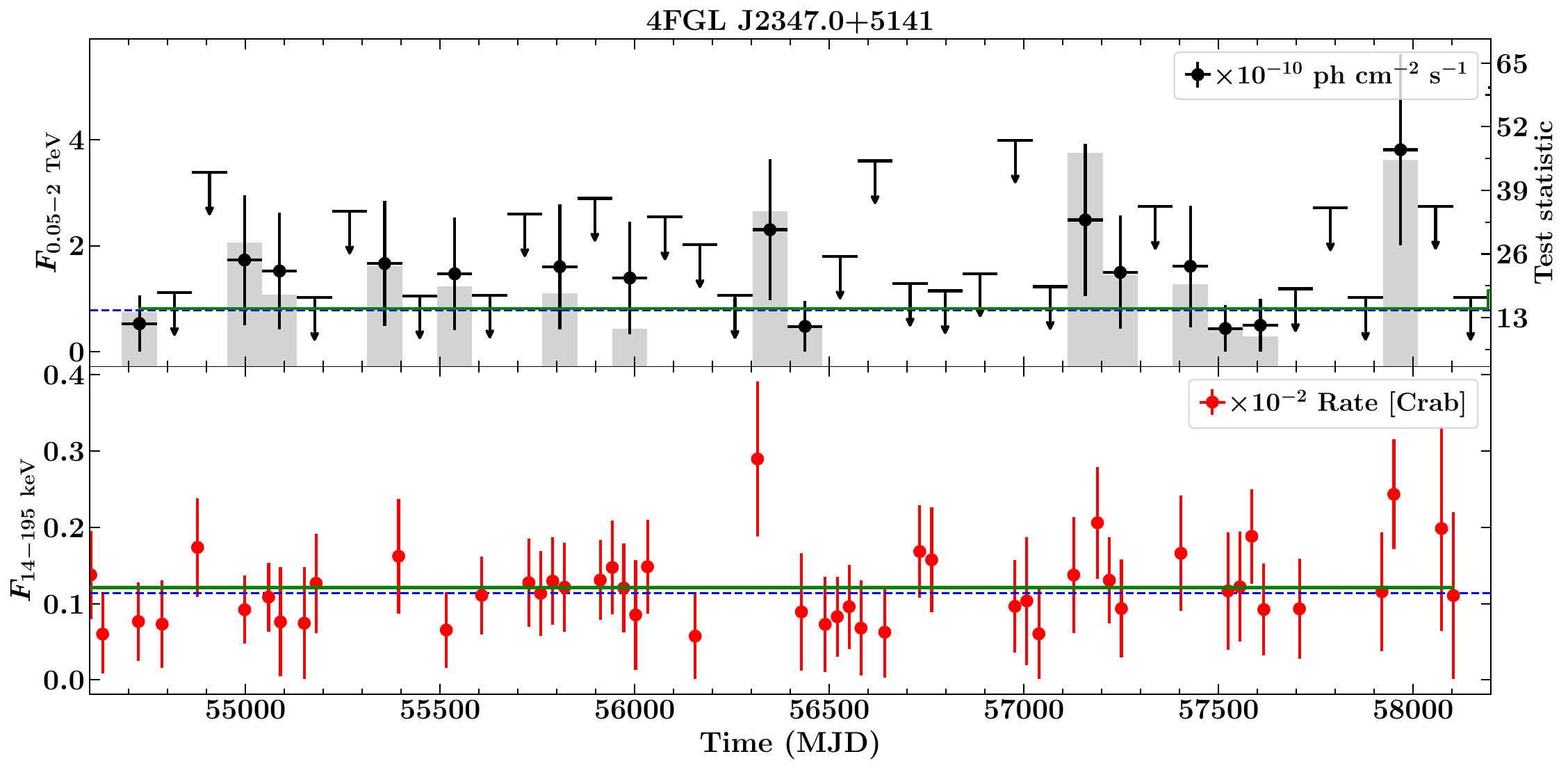}
}
\caption{Figure 2 (continued)}
\end{figure}

\section{Results}\label{sec4}
The light curves for both TeV and GeV bands are shown in Figure~\ref{fig:1}. In the low-energy GeV band, most of the sources exhibited strong flux variability, however, only a small fraction appeared variable at very high frequencies. The extent of temporal flux variations was quantified by estimating the variability test statistic (\tsv) following a maximum-likelihood approach \citep[cf.][]{2012ApJS..199...31N,2017ApJ...841..100A}:

\begin{equation}
\label{eq:TSvar}
 \mathrm{TS}_{\mathrm{var}} =  2 \sum_i{ \left( \ln{\mathcal{L}_i(F_i)} - 
\ln{\mathcal{L}_i(F_{\mathrm{avg}})} \right) }
\end{equation}
where  $\mathcal{L}_i(F_i)$ is the likelihood with $F_i$ being the best-fit flux value for time bin $i$. The parameter $\mathcal{L}_i(F_{\mathrm{avg}})$, is the likelihood with $F_{\mathrm{avg}}$ being is the best-fit flux value for the full time period derived from the likelihood fitting. The parameter \tsv~is distributed as $\chi^2$ with 61 degrees of freedom if the null hypothesis is correct, and a value of TS$_{\rm var}>80.2$ or $>89.6$ corresponds to a source being variable at $>$95\% or $>$99\% confidence level, respectively. The \tsv~values estimated for all sources are reported in Table~\ref{tab:basic_info}. A blazar was considered variable (V) for $>$99\% confidence level and probably variable (PV) if it had a value between 95\% and 99\%. The remaining sources were considered non-variable (NV). This exercise led to the identification of 5 objects showing significant flux variability in the TeV band, including Mkn~421, which is the most variable source in the sample. There are 8 blazars whose TeV light curves hint for the existence of flux variability, albeit at a lower confidence level. Remaining 16 objects did not show any significant variability. Even though several blazars do not show strong VHE flux variations, likely due to low photon statistics, interesting variability patterns emerge when comparing the light curves in the TeV and GeV energy bands. To quantify the observed patterns, a time-lag analysis was carried out using the \textit{z-transformed discrete correlation function} (\dcf) technique and the uncertainties were estimated using a likelihood method \citep[][]{1997ASSL..218..163A,2013arXiv1302.1508A}. The \dcf~method provides a conservative estimate for the cross-correlation as a function of time-lag between unevenly sampled light curves. Furthermore, blazars often exhibit erratic flux variations with the same object some time showing correlated flux changes and uncorrelated variability during other epochs \citep[e.g., 3C 279,][]{2015ApJ...807...79H,2015ApJ...803...15P,2016ApJ...817...61P}. Additionally, only flux upper limits could be derived during several epochs for various objects under consideration and that may somewhat bias the result as those upper limits would cover mainly the low states of the source. Therefore, rather than running on the full dataset, \dcf~analysis was carried out separately on specific epochs. The time periods were selected in such a manner to cover the interesting activity epochs while maximizing the number of data points. The results are discussed in the next section.

The Bayesian block representation of the time series data allows to identify the periods of different activity states.  An advantage of this technique is that it is able to identify significant changes in light curve independently of variations in exposure or gaps \citep[][]{2013ApJ...764..167S,2022icrc.confE.868W}. Therefore, the best-fit piece wise constant representations of the \fermi-LAT data were derived using this method. The false-alarm probability was chosen as 25\% and 5\% for the TeV band and all other (GeV band and hard X-ray) light curves, respectively. For 22 out of 29 objects, this exercise resulted in the identification of blocks of different TeV band flux states. The results are shown in Figure~\ref{fig:1}.

In the conventional leptonic radiative models, the VHE radiation observed from high-synchrotron peaked BL Lac objects is produced by the highest energy electrons via inverse Compton scattering of synchrotron photons \citep[e.g.,][]{2010MNRAS.401.1570T}. At lower frequencies, the same electron population is expected to radiate synchrotron emission usually lying in the hard X-ray band \citep[$>$10 keV,][]{2018MNRAS.477.4257C,2019MNRAS.486.1741F}. Therefore, it is imperative to compare the TeV band flux variations with that observed at hard X-rays. For this purpose, Crab weighted, monthly binned light curves of 11 \swift-Burst Alert Telescope (BAT) detected objects were retrieved from the 157-month \swift-BAT hard X-ray survey website\footnote{\url{https://swift.gsfc.nasa.gov/results/bs157mon/}}. The \swift-BAT data overlaps with the \fermi-LAT observations in the time period MJD 54683$-$58103, i.e., up to 2017 December 16.  For the analysis, only those hard X-ray data points were considered that have the signal-to-noise ratio $>$1. The light curves for both TeV and hard X-ray bands are shown in Figure~\ref{fig:2}.

\section{Notes on individual objects}\label{sec5}
{\it 1ES 0033+595}: The TeV band light curve of this Major Atmospheric Gamma Imaging Cherenkov (MAGIC) telescopes detected source \citep[][]{2015MNRAS.446..217A} revealed a flaring activity in the time period MJD 56392$-$57562. Interestingly, the pattern of the GeV flux variations suggests the flare to peak at a later time than the TeV flare. However, the results of the \dcf~analysis indicated the absence of a significant lead/lag due to large uncertainties (Table~\ref{tab:dcf}). The Bayesian block analysis suggested a high flux state overlapping in both TeV and GeV bands. The source is overall non-variable in the TeV band as per the \tsv~calculation. Furthermore, a comparison of the TeV and 14$-$195 keV light curves revealed the source to be in an elevated hard X-ray activity state at the time of the TeV flare (Figure~\ref{fig:2}).

{\it 3C 66A}: This well-known Very Energetic Radiation Imaging Telescope Array System (VERITAS) detected blazar was in a high GeV activity state during the first year of the \fermi-LAT operation \citep[][]{2011ApJ...726...43A}. The high activity was also reflected in the TeV band light curve \citep[Figure~\ref{fig:1};][]{2011ApJ...726...58A}. The \dcf~analysis suggested the contemporaneous flux enhancements in both TeV and GeV bands (Table~\ref{tab:dcf}). Later, the object went into a low activity state but was frequently detected in the TeV band at a flux level comparable to its $\sim$15 years averaged value. The source can be considered moderately variable based on the estimated \tsv~value. The Bayesian block analysis carried out on the TeV band light curve hinted for the presence of variable flux activity states.

{\it PKS 0447$-$439}: The TeV emission from this high-synchrotron peaked blazar was first detected with the High Energy Stereoscopic System (H.E.S.S.) during 2009-2010 \citep[][]{2013A&A...552A.118H} consistent with the low-activity detection in the TeV band light curve (Figure~\ref{fig:1}). The source has shown moderate variability in the TeV band. It exhibited its brightest TeV flare in the time bin MJD 59002$-$59092. This activity was also reflected in the GeV band light curve. Interestingly, a GeV flare of similar amplitude was observed from this object earlier in the time bin MJD 57652$-$57742, however, the source was detected at its $\sim$15-years averaged flux level in the TeV band. To ascertain this, the \dcf~analysis was employed on two time periods: MJD 56752$-$57922 and MJD 58282$-$59272. The \dcf~coefficient for the first period was found to be 0.29$^{+0.42}_{-0.50}$ indicating the two events to be only marginally correlated. For the second period, TeV and GeV flux variations were strongly correlated (\dcf~coefficient = 0.89$^{+0.09}_{-0.14}$) with no significant lead/lag. Moreover, the Bayesian block analysis revealed the existence of variable flux activity states and the source is found as mildly variable based on the estimated \tsv~value.

\begin{table}
\caption{The $z$-DCF analysis results. The column information are as follows: (1) 4FGL name of the source; (2) time period considered for the $z$-DCF analysis, in MJD; (3) the $z$-DCF coefficient value; (4) the time lag between the two light curves in days with a positive value indicates the GeV band flux to lag behind TeV band flux, in days.}\label{tab:dcf}
\begin{center}
\begin{tabular}{llll}
\hline
4FGL name  & Time period & $z$-DCF coeff. & Time lag\\
(1) & (2) & (3) & (4) \\ 
\hline
  & GeV$-$TeV bands & & \\
J0035.9+5950    & 56032$-$57562 & 0.21$^{+0.25}_{-0.26}$ & 90$^{+228}_{-117}$ \\
J0222.6+4302    & 54683$-$55762 & 0.85$^{+0.11}_{-0.17}$ & 0$^{+293}_{-50}$ \\
J0449.4$-$4350 & 56752$-$57922 & 0.29$^{+0.42}_{-0.50}$ & 450$^{+443}_{-550}$ \\
                                & 58282$-$59272 & 0.89$^{+0.09}_{-0.14}$ & 0$^{+265}_{-234}$ \\
J0507.9+6737      & 54683$-$55582 & $-$0.13$^{+0.59}_{-0.54}$ & 360$^{+400}_{-306}$ \\
J1104.4+3812       & 56032$-$57382 & 0.35$^{+0.21}_{-0.26}$ & 25$^{+195}_{-309}$ \\
J1427.0+2348     & 55357$-$56707 & 0.33$^{+0.34}_{-0.40}$ & 0$^{+243}_{-298}$ \\
J1555.7+1111        & 58462$-$59542 & 0.31$^{+0.40}_{-0.42}$ & $-$90$^{+149}_{-150}$ \\
J1653.8+3945     & 55762$-$57562 & 0.27$^{+0.23}_{-0.25}$ & $-$90$^{+95}_{-134}$ \\
J2202.7+4216    & 58822$-$60082 & 0.87$^{+0.09}_{-0.13}$ & 0$^{+42}_{-43}$ \\
\hline
  & keV$-$TeV bands & & \\
J0507.9+6737   & 54683$-$55582 & 0.65$^{+0.29}_{-0.51}$ & $-$165$^{+269}_{-378}$ \\
J1104.4+3812   & 56032$-$57382 & 0.17$^{+0.41}_{-0.36}$ & $-$451$^{+251}_{-464}$ \\
J1653.8+3945   & 55762$-$57562 & 0.67$^{+0.13}_{-0.21}$ & $-$40$^{+73}_{-109}$ \\

\hline
\end{tabular}
\end{center}
\end{table}

{\it 1ES 0502+675}: A $\sim$2.5-year long TeV flaring event was observed by the \fermi-LAT during the initial years of its operation, and the source was also detected with VERITAS in this time-period \citep[Figure~\ref{fig:1};][]{2009ATel.2301....1O}. Due to large uncertainties in both TeV and GeV bands data, however, the results of the \dcf~analysis were inconclusive. The source went into quiescence after this flare and was sporadically detected in the TeV band at a flux level comparable to its $\sim$15 years averaged value as also confirmed by the Bayesian block analysis. Overall, significant TeV flux variability at $>$99\% confidence level was detected from this blazar. The source appeared to be bright in hard X-rays during the elevated TeV activity state which is confirmed by the \dcf~analysis. There are hints for the X-ray flare leading the TeV event, however, it cannot be claimed due sparseness of the data points and large uncertainties in the derived parameters (Table~\ref{tab:dcf}).

{\it TXS 0518+211}: This high-synchrotron peaked BL Lac object was detected with VERITAS during MJD 55126$-$55212 \citep[][]{2013ApJ...776...69A}. A flaring activity in the TeV band was identified in the time period MJD 56302$-$56752 which was reflected in the GeV band and also followed by the Cherenkov telescopes \citep[][]{2022ApJ...932..129A}. Later, the source was detected in the TeV band on a few occasions and found to exhibit moderate flux variations as quantified by the \tsv~analysis.

{\it RX J0543.9$-$5532}: \citet[][]{2014MNRAS.445.4345B} reported the first VHE detection of this source using the \fermi-LAT data above 100 GeV though it was not detected with the Cherenkov telescopes \citep[][]{2014A&A...564A...9H}. The GeV and TeV band light curves indicated that the object was in quiescence most of the time since the \fermi-LAT operation began in 2008. According to the Bayesian block analysis, the enhanced GeV band activity was identified starting with the time bin MJD 58642$-$58732 and the source was also detected in the TeV band. However, no significant TeV flux variability was identified as revealed by the \tsv~analysis.

{\it RX J0648.7+1516}: This blazar was detected only on a few occasions in the TeV band during the $\sim$15 years of the \fermi-LAT operation and can be considered non-variable from the estimated \tsv. The GeV light curve of this object also does not exhibit any significant flux variability. 

{\it 1ES 0647+250}: This object is one of the few blazars detected at TeV energies by MAGIC telescopes even during non-flaring activity states \citep[Figure~\ref{fig:1};][]{2023A&A...670A..49M}. No significant flux variations were noticed in the TeV band light curve  which was also confirmed by the derived \tsv~value.

{\it S5 0716+71}: This \gm-ray bright blazar has exhibited exceptional GeV flaring activities and is also detected at TeV energies by the MAGIC telescopes \citep[][]{2009ApJ...704L.129A,2018A&A...619A..45M,2020ApJ...904...67G}. The large amplitude flux variability can also be seen in the GeV band light curve, whereas, the flux variations observed in the TeV band data were minor. The largest TeV band flare was observed in the time period MJD 56032$-$56212 which was also reflected at lower energies. The elevated TeV flux activity was also detected by Cherenkov telescopes on several occasions which are consistent with the TeV band detections \citep[Figure~\ref{fig:1};][]{2017ATel11100....1M,2018A&A...619A..45M}. Overall, the estimated \tsv~has indicated that the source is partially variable. This object is also detected with \swift-BAT and a comparison of the TeV and hard X-ray light curves revealed the lack of hard X-ray counterpart during the TeV flare (MJD 56032$-$56212). Though it was not possible to quantify the lack of correlation due to sparseness of the data points, this observation can be understood from the SED behavior of the source. It is an intermediate synchrotron peaked BL Lac object whose `valley' between the two broad humps lies in the X-ray band \citep[cf.][]{2018A&A...619A..45M}. In such a scenario, the hard X-ray ($>$10 keV) radiation is primarily dominated by the inverse Compton emission produced by the low-energy electron population, thus, less variable than the VHE emission produced by the highest energy electrons.

{\it 1ES 0806+524}: This object was detected in the TeV band on several occasions at a flux level consistent with its $\sim$15 years averaged value. The \tsv~analysis has revealed it to be non-variable and the Bayesian block analysis has also not found any significant variable flux states. The GeV band light curve highlighted a flaring activity in the time period MJD 55582$-$55852 in which the TeV band flux shows an enhancement as well. The source was also detected with MAGIC telescopes during this flaring episode \citep[Figure~\ref{fig:1};][]{2015MNRAS.451..739A}.

{\it 1H 1013+498}: This blazar was detected with the MAGIC telescopes coincident with the observation of an optical outburst \citep[][]{2007ApJ...667L..21A}. From the \fermi-LAT data analysis, only a moderate level of flux variability was noticed in both the GeV and TeV band light curves, also confirmed by the Bayesian block routine. The source is non-variable according to the estimated \tsv~value. A comparison of the TeV and GeV bands light curves revealed that the elevated flux activities noticed in the former were also reflected in the latter. Furthermore, the source was detected with Cherenkov telescopes during these epochs \citep[][]{2016A&A...590A..24A,2016A&A...591A..10A}.

{\it GB6 J1037+5711}: The TeV light curve of this object does not show any significant flux variability as confirmed by the \tsv~value. It was detected only on a few occasions in the TeV band and the Bayesian block analysis did not reveal any significant variable flux states. The High Altitude Water Cherenkov Observatory reported the detection of a TeV transient event from this blazar during MJD 58298$-$58300 \citep[][]{2018ATel11806....1W}. This epoch lies in the time bin MJD 58282$-$58372, in which GB6 J1037+5711 was significantly detected in the TeV band (Figure~\ref{fig:1}).

{\it Mkn 421}: This object is probably the brightest and the most studied TeV blazar to date \citep[cf.][]{2011ApJ...736..131A}. It was detected in the TeV band throughout the $\sim$15 years of the \fermi-LAT operation and also by several Cherenkov telescopes \citep[Figure~\ref{fig:1};][]{2016ApJS..222....6B,2019ATel12683....1G,2020ApJ...890...97A,2021A&A...647A..88A}. The pattern of the flux variability in both the GeV and TeV bands appears similar, with the high/low TeV activity states coinciding with that observed in the GeV band. However, the object also exhibited complex variability behavior on a few occasions. For example, the high activity noticed in the TeV band in the time bin MJD 56122$-$56212 was also seen in the GeV band, and the estimated GeV flux was found to be the highest observed from the source during the first $\sim$15 years of the \fermi-LAT operation. On the other hand, the GeV flux corresponding to a TeV flare of similar amplitude, observed in the time bin MJD 56482$-$56572, was considerably lower than that observed during the previous TeV flare described above. The \dcf~analysis also indicates the existence of a mild correlation (Table~\ref{fig:2}). The observation of such uncorrelated or weakly correlated GeV$-$TeV flux variations from Mkn 421 are not uncommon \citep[e.g.,][]{2016ApJS..222....6B,2020ApJS..248...29A}. Interestingly, during the both TeV flaring events, Mkn 421 was too close to the Sun to be followed up by the multi-wavelength observatories implying that two of the brightest TeV outbursts detected by the \fermi-LAT could not be followed up. This finding also highlights the importance of an all-sky surveying \gm-ray mission to capture the extraordinary flaring episodes, some of which, otherwise, may remain unexplored due to practical constraints for other facilities. Interestingly, a comparison of the TeV and hard X-ray light curves reveal the source to be in quiescence in the 14$-$195 keV band at the time of the TeV flare observed in the time bin MJD 56122$-$56212 (Figure~\ref{fig:2}). This uncorrelated variability pattern was also confirmed with the \dcf~analysis (Table~\ref{tab:dcf}). This is probably a surprising result since the VHE and X-ray observations are often strongly correlated \citep[e.g.,][]{2019MNRAS.484.2944G}.

{\it RBS 0970 and Mkn 180}: The TeV band light curves of these two blazars do not exhibit any significant flux variability and detected only on a few occasions. Furthermore, the GeV flux variations were modest, as seen in their light curves. The \tsv~analysis revealed both of them to be non-variable.

{\it 1ES 1215+303}: The long-term VHE flux variability behavior of this source was recently studied by \citet[][]{2020ApJ...891..170V} using VERITAS observations and significant flux variations were detected \citep[see also][]{2012A&A...544A.142A}. The TeV band light curve has reveled several \gm-ray flaring episodes which were also found by the Bayesian block algorithm. The high activity was reflected in the GeV band light curve. Based on the estimated \tsv~parameter, this object can be considered partially variable.

{\it PG 1218+304}: The TeV band light curve of this high-synchrotron peaked blazar showed a low-amplitude \gm-ray flare in the first year of the \fermi-LAT operation which coincided with the detection by VERITAS and observation of day-scale flux variability \citep[Figure~\ref{fig:1};][]{2010ApJ...709L.163A}. The source was detected several times displaying a moderate level of TeV flux variability which is confirmed by the derived \tsv~value (Table~\ref{tab:basic_info}). The Bayesian block analysis also revealed the presence of variable flux states. A comparison of the TeV and hard X-ray light curves indicated that the low-level flux variability is present in both bands.

{\it PKS 1424+240}: At the redshift of $z=0.60$ \citep[][]{2017ApJ...837..144P}, this is the most distant blazar among the sources considered in this work. This blazar was significantly detected in several time bins of the TeV band light curve and also by the VERITAS and MAGIC telescopes \citep[][]{2010ApJ...708L.100A,2014A&A...567A.135A,2014ApJ...785L..16A}.  The TeV flux variations were accompanied by that observed in the GeV band light curve which was confirmed by the overlapping Bayesian blocks. The results of the \dcf~analysis carried out during the high-activity period (MJD 55357$-$56707) also supported this finding though a strong claim cannot be made due to the large uncertainties in the \dcf~coefficient (Table~\ref{tab:dcf}). The estimated \tsv~parameter remained below the variability threshold thereby rendering this object as non-variable.

{\it PKS 1440$-$389}: This blazar has not exhibited any significant flux variability in the TeV band. It was sporadically detected throughout the $\sim$15 years of the \fermi-LAT operation and and flux enhancement was seen mainly in the later half of the covered time period. The GeV band flux showed a similar behavior and the brightest GeV flux identified in the time bin MJD 59002$-$59092 had a TeV band counterpart. The Bayesian block analysis has found variable flux states. This source was detected with the H.E.S.S. during a low activity state \citep[][]{2020MNRAS.494.5590A}.

{\it PG 1553+113}: This bright VHE emitting blazar has exhibited quasi-periodic variability in the \gm-ray band \citep[][]{2015ApJ...813L..41A}. It was detected almost throughout all the analyzed periods and has been detected with several Cherenkov telescopes \citep[Figure~\ref{fig:1};][]{2012ApJ...748...46A,2015ApJ...799....7A,2015ApJ...802...65A,2017A&A...600A..89H,2019ICRC...36..689G}. Based on the \tsv~value, this object is partially variable in the TeV band. Several episodes of moderate amplitude, i.e., a factor of $\sim$2 with respect to the $\sim$15 years averaged flux, flaring activity have been noticed which were also reflected in the GeV band light curve. In the time period MJD 58462$-$59542, two similar amplitude TeV flaring events peaking in the time bins MJD 58642$-$58732 and MJD 59272$-$59362 were identified. Though the GeV band light curve showed flux enhancement at both epochs, the amplitude of flux variations were different with the second flare having higher GeV flux which is also confirmed by the Bayesian block analysis. The source was detected with MAGIC telescopes during this flare \citep[][]{2021ATel14520....1B}. The \dcf~analysis hinted an overall positive correlation though the uncertainties in the estimated \dcf~coefficient are large (Table~\ref{tab:dcf}).

{\it Mkn 501}: This well-studied, bright TeV blazar has exhibited violent flux variability across the electromagnetic spectrum \citep[e.g.,][]{2019ApJ...870...93A}. The TeV band light curve of this source has revealed several episodes of VHE outbursts, with some of them also being detected with various Cherenkov telescopes \citep[cf.][]{2018A&A...620A.181A,2019ApJ...870...93A,2023ApJS..266...37A}. The estimated \tsv~parameter suggests the source to be variable in the TeV band. Furthermore, it remained mostly in quiescence during the later half of the \fermi-LAT observation period. During the periods of high TeV activity, e.g., in the time interval MJD 55762$-$57562, the GeV band flux showed similar variations within uncertainties, which was also confirmed by the \dcf~analysis (Table~\ref{tab:dcf}). This source is detected with the \swift-BAT and a comparison of the TeV band and hard X-ray light curves indicated contemporaneous flux variations in both energy bands which was supported by the \dcf~analysis.

{\it I Zw 187}: This blazar was significantly detected only on a few occasions in the TeV band light curve and has also been detected with Cherenkov telescopes \citep[Figure~\ref{fig:1};][]{2014A&A...563A..90A,2015ApJ...808..110A}. A high TeV band activity was noticed in the time period MJD 56572$-$57112 and was found to have coincided with an elevated GeV flaring episode as also confirmed by the Bayesian block analysis. The source is non-variable in the TeV band as per the computed \tsv~value (Table~\ref{tab:basic_info}).

{\it 1H 1914$-$194}: This object was detected in the TeV band only on a few occasions, and the \tsv~parameter revealed it to be non-variable. The GeV band light curve shows only a moderate level of flux variability. The Bayesian block analysis did not find any significant variable flux states.

{\it 1ES 1959+650}: This source is a bright VHE emitting blazer \citep[e.g.,][]{2013ApJ...775....3A}, and the detection of an orphan TeV flare was also reported \citep[][]{2004ApJ...601..151K}. This is one of the few sources studied in this work that were consistently detected in the TeV band throughout the $\sim$15 years of the \fermi-LAT operation. The TeV band light curve of this blazar has indicated significant flux variability, which was also accompanied by the flux variations seen in the GeV band light curve. In particular, a $\sim$4 years long ($\sim$MJD 57022$-$58462) \gm-ray flare was identified in both the TeV and GeV band light curves which was confirmed by the Bayesian block analysis. The multiwavelength observations close in time to the peak of this \gm-ray flare have been reported by \citet[][]{2020A&A...638A..14M}. 
In the time period covered by the \fermi-LAT and \swift-BAT, similar flux variations were observed in TeV and hard X-ray bands light curves.

{\it PKS 2005$-$489}: This blazar was in an elevated activity state during the first year of the \fermi-LAT operation as revealed by its TeV, GeV, and hard X-ray light curves. The source was also detected with the H.E.S.S. observatory in this time period \citep[Figure~\ref{fig:1};[]{2011A&A...533A.110H}. After that, the object went into quiescence for the rest of the observing period, exhibiting no significant variability and was sporadically detected at TeV energies. Overall, it is non-variable in the TeV band based on the estimated \tsv~value.

{\it RGB J2056+496}: This object was occasionally detected in the TeV band at a flux level consistent with the $\sim$15 years averaged value. The calculated \tsv~parameter suggested this blazar to be non-variable. This object was also detected with the VERITAS \citep[][]{2016ATel.9721....1M}. The hard X-ray band light curve of this object showed an enhanced flux state around MJD 57524 (Figure~\ref{fig:2}), however, the presence/absence of the TeV counterpart could not be quantified due to sparseness of the data points.

{\it PKS 2155$-$304}: This bright VHE-detected blazar has exhibited significant flux variability at GeV energies and has been studied from several Cherenkov telescopes \citep[Figure~\ref{fig:1}][]{2017A&A...598A..39H,2017A&A...600A..89H,2019hepr.confE..27W}. The detection of the rapid, $\sim$minute-scale, TeV flux variability was also reported \citep[][]{2007ApJ...664L..71A}. In the TeV band, it was found to be partially variable based on the derived \tsv~value.  As seen in its TeV band light curve, it was almost regularly detected throughout the time period considered in this work. The elevated TeV activity states were also reflected in the GeV band.

{\it BL Lacertae}: This eponymous blazar has exhibited several \gm-ray flares since the \fermi-LAT observations began in 2008, with the most prominent one observed during 2020-2021 \citep[cf.][]{2022A&A...668A.152P}. In the TeV band, the source remained in quiescence for most of the observing run prior to the 2020-2021 flaring episode though a rapid TeV flare was detected in 2015 June \citep[][]{2019A&A...623A.175M}. It was regularly detected in the TeV band during MJD 58822$-$60082, i.e., the period of the high activity, and exhibited huge TeV flares, consistent with its detection with MAGIC telescopes in this period\citep[e.g.,][]{2020ATel13963....1B,2021ATel14826....1B}. The GeV band light curve also revealed large amplitude flares contemporaneous with the elevated TeV activity which was confirmed with the \dcf~analysis (Table~\ref{tab:dcf}). During the time period common to \fermi-LAT and \swift-BAT, the source remained in quiescence in the hard X-ray and TeV bands.

{\it 1ES 2344+514}: This source was sporadically detected in the TeV band light curve and no significant flux variability was noticed (Table~\ref{tab:basic_info}). In the GeV band also, it has exhibited a moderate level of flux variations. This source has been detected several times with various Cherenkov telescopes \citep[][]{2013A&A...556A..67A,2017MNRAS.471.2117A,2020MNRAS.496.3912M}. A comparison of the TeV and hard X-ray bands light curves indicated no significant flux variability.

\section{Summary}\label{sec6}

This work presents the first results of the 0.05$-$2 TeV LAT data analysis to explore the long-term VHE flux variations of bright \fermi-LAT detected blazars.  Among the sample of 29 objects, 5 sources were found to exhibit statistically significant TeV band flux variability, whereas, 8 of them have shown moderate levels of VHE flux variations. The remaining 16 blazars were found to be non-variable, however, some of them have exhibited minor TeV flaring activities in the time period covered in this work. Overall, these results lead to the conclusion that blazars exhibit long-term flux variations at TeV energies. Another important finding of this work is the observation of complex variability patterns when comparing the TeV and GeV band light curves. In some cases, e.g., Mkn 421, the TeV flaring events had weak/no GeV counterpart and the GeV emission peaked before/after the peak of the TeV flaring activity in some other sources. Admittedly, a strong claim could not be made due to sparse data points in the TeV band light curves for the majority of sources. Furthermore, in the standard one-zone leptonic emission model, the VHE and hard X-ray radiations produced by the jet are expected to be originated from the same electron population of the highest energies. Indeed, the TeV band variability patterns observed from the \swift-BAT detected blazars present in our sample were often found to be similar to that identified in the 14$-$195 keV, monthly-binned, light curves indicating co-spatiality of the produced emission. Interestingly, the TeV band light curve of the bright blazar Mkn~421 revealed a flare that had no hard X-ray counterpart.
These findings provide clues about the complex radiative mechanisms operating in blazar jets and are aligned with the results obtained in several recent multiwavelength campaigns focused on TeV blazars where complex variability and correlation patterns have been found \citep[cf.][]{2020ApJS..248...29A,2021A&A...655A..89M}. In addition, further temporal and spectral investigation of these flaring episodes may also enable one to probe the radiative processes and possible interaction of the VHE emission with the extragalactic background light \citep[e.g.,][]{2015ApJ...813L..34D,2020NatAs...4..124B}. All in all, the findings presented in this work should be considered as the first step to unravel the long-term VHE variability behavior of blazars, which will be explored in greater detail with the upcoming Cherenkov Telescope Array due to its unprecedented sensitivity.

\acknowledgements
The author thanks the journal referee for constructive criticism. This research has made use of NASA’s Astrophysics Data System Bibliographic Services. The use of the \fermi-LAT data provided by the Fermi Science Support Center is gratefully acknowledged.

\software{fermiPy \citep{2017arXiv170709551W}}.

\bibliographystyle{aasjournal}

\end{document}